\documentclass{iopart}
\usepackage{iopams}
\usepackage{amssymb}
\usepackage{bm}
\usepackage{graphics}
\usepackage{graphicx}
\usepackage{color}
\eqnobysec

\begin{document}
\topical[The role of the Riemann-Silberstein vector in classical and quantum theories of electromagnetism]{The role of the Riemann-Silberstein vector in\\ classical and quantum theories of electromagnetism}

\author{Iwo Bialynicki-Birula$^1$ and Zofia Bialynicka-Birula$^2$}
\address{$^1$ Center for Theoretical Physics, Polish Academy of Sciences,\\Al. Lotnik\'ow 32/46, 02-668 Warsaw, Poland\\ $^2$ Institute of Physics, Polish Academy of Sciences,\\Al. Lotnik\'ow 32/46, 02-668 Warsaw, Poland}
\ead{birula@cft.edu.pl}

\begin{abstract}
It is shown that the use of the Riemann-Silberstein (RS) vector greatly simplifies the description of the electromagnetic field both in the classical domain and in the quantum domain. In this review we describe many specific examples where this vector enables one to significantly shorten the derivations and make them more transparent. We also argue why the RS vector may be considered as the best possible choice for the photon wave function.
\end{abstract}
\submitto{Journal of Physics A}

\tableofcontents
\pagestyle{myheadings}
\markright{Topical Review}
\vspace{1.0cm}

\section{Introduction}

This review is devoted to various applications in classical and in quantum theory of the following complex combination of the electromagnetic field vectors:
\begin{eqnarray}\label{rs}
{\bm F} = \frac{\bm D}{\sqrt{2\epsilon}}+\rmi\frac{\bm B}{\sqrt{2\mu}},
\end{eqnarray}
where $\epsilon$ is the dielectric constant and $\mu$ is the magnetic permeability. The denominators in the definition of the  RS vector are needed to make the dimensions of both terms equal. The RS vector is most useful when dealing with pure radiation field in empty space or in a homogeneous and static medium where $\epsilon$ and $\mu$ are constant. In this review we restrict ourselves most of the time to this case.

\begin{figure}
\centering
\includegraphics[scale=0.9]{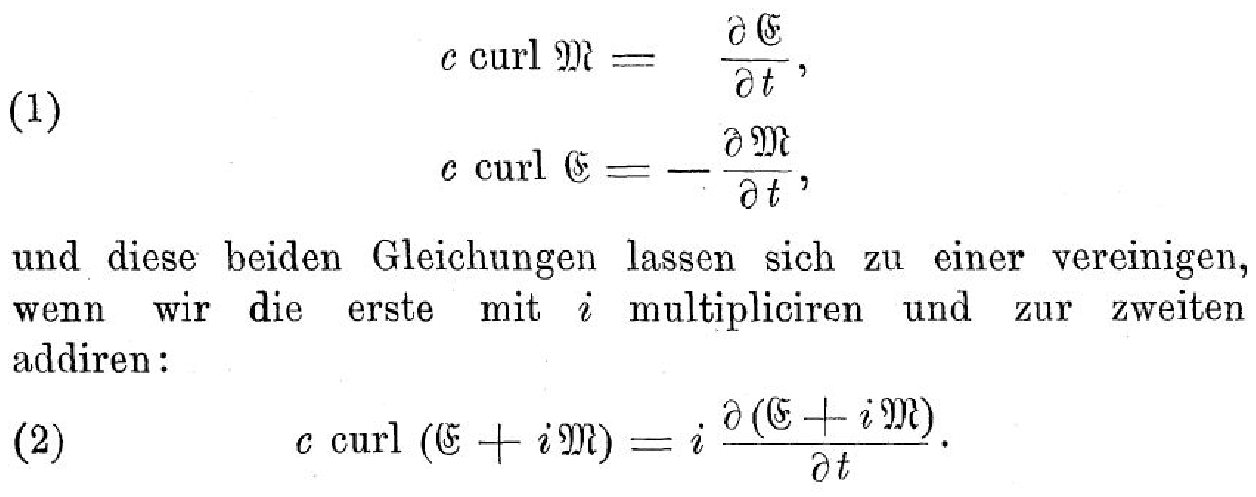}
\caption{The first appearance in print of the Riemann-Silberstein vector}\label{fig1}
\end{figure}

The RS vector has been studied in some detail for the first time in 1907 by Ludwik Silberstein \cite{sil,sil1}. In the second paper Silberstein wrote that this vector made its first appearance (see Fig.~\ref{fig1}) in the lectures on partial differential equations by Bernhard Riemann edited and published in 1901 by Heinrich Weber \cite{web}. Later Silberstein incorporated this vector into the quaternionic formulation of the relativity theory \cite{sil2}. The name the Riemann-Silberstein (RS) vector was introduced by us in \cite{pwf} and this name is now generally accepted.

This review is divided into two parts: classical theory of electromagnetism and quantum electrodynamics. Many results presented here are not original---most of them are taken from our earlier publications [5--23]. We shall show that the RS vector offers a uniform description linking the wave aspect and the corpuscular aspect of electromagnetism. This link has also been emphasized in a recent article by Smith and Raymer~\cite{sr}. We must also mention here very extensive reviews by Keller~\cite{kel} and Saari~\cite{saari} where the RS vector is used to describe quantum mechanics of photons with special emphasis on photon localization.

The Maxwell equations in the absence of sources read:
\numparts
\begin{eqnarray}
\partial_t{\bm D}({\bm r},t)&={\bm\nabla}\times{\bm H}({\bm r},t),\label{max0a}\\
\partial_t{\bm B}({\bm r},t)&=-{\bm\nabla}\times{\bm E}({\bm r},t),\label{max0b}\\
{\bm\nabla}\cdot{\bm D}({\bm r},t)&=0,\label{max0c}\\
{\bm\nabla}\cdot{\bm B}({\bm r},t)&=0.\label{max0d}
\end{eqnarray}
\endnumparts
By adding the first equation multiplied by $1/\sqrt{2\epsilon}$ to the second equation multiplied by $\rmi/\sqrt{2\mu}$ and doing the same for the third and fourth equation one obtains two complex equations
\numparts
\begin{eqnarray}
\rmi\partial_t{\bm F}({\bm r},t)&={\bm\nabla}\times{\bm G}({\bm r},t),\label{maxca}\\
{\bm\nabla}\cdot{\bm F}({\bm r},t)&=0,\label{maxcb}
\end{eqnarray}
\endnumparts
where the second complex vector is defined as:
\begin{eqnarray}\label{grs}
{\bm G} = \frac{\bm E}{\sqrt{2\mu}}+\rmi\frac{\bm H}{\sqrt{2\epsilon}}.
\end{eqnarray}
In vacuum (or in a homogeneous and static medium), due the simple form of the constitutive relations,
\begin{eqnarray}\label{cr0}
{\bm H}={\bm B}/\mu,\quad {\bm E}={\bm D}/\epsilon,
\end{eqnarray}
the vector ${\bm G}$ is a multiple of ${\bm F}$ (${\bm G}=c{\bm F})$. In this case the four Maxwell equations are equivalent to the following equations for the complex RS vector:
\begin{eqnarray}
\rmi\partial_t{\bm F}({\bm r},t)&= c{\bm\nabla}\times{\bm F}({\bm r},t),\label{maxa}\\
{\bm\nabla}\cdot{\bm F}({\bm r},t)&=0.\label{maxb}
\end{eqnarray}
Thus, with the use of the RS vector, one reduces the number of equations by a factor of two at a price of working with a complex field. Note that when $\epsilon$ and $\mu$ vary in space or in time the reduction to two equations does not work \cite{pwf}.

The use of a complex vector to describe the electromagnetic field is not a complication since it is often customary, as a matter of convenience, to seek complex solutions of Maxwell equations. The imaginary part is rejected at the end as nonphysical. We are not so wasteful when we are working with the RS vector since both parts are physically meaningful. The real part of the RS vector describes the electric field and the imaginary part describes the magnetic field. The economy, however, is not the only reason to use the RS vector.

The usefulness of the RS vector stems from several factors.
\begin{itemize}
\item{The use of such a complex representation of the electromagnetic field greatly simplifies the derivation of various solutions of Maxwell equations. This has been exploited already long time ago by Bateman \cite{bateman}.}
\item{The complexification of the electromagnetic field allows for generation of new solutions from the old ones.}
\item{The use of the RS vector underscores the role of the Hertz and Debye superpotentials.}
\item{The RS vector is as close to the photon wave function in coordinate space as possible, taking into account the limitations imposed by the lack of the position operator for photons.}
\item{The quantization of the electromagnetic field is more transparent when the field operators are combined into the RS field operator.}
\end{itemize}
In this review we will illustrate the merits of the RS vector with many specific examples.

\section{Basic properties of the RS vector}

In the absence of sources the splitting of the electromagnetic field into the electric and magnetic part is not absolute because in this case Maxwell equations are invariant under the {\em duality rotation} of the electromagnetic field (\cite{dt},\cite{jack} p. 274),
\numparts
\begin{eqnarray}\label{dual}
{\bm E}' = {\bm E}\cos\varphi-c{\bm B}\sin\varphi,\\
c{\bm B}' = c{\bm B}\cos\varphi+{\bm E}\sin\varphi.
\end{eqnarray}
\endnumparts
The duality rotation expressed in terms of the RS vector means the multiplication of this vector by a constant phase factor,
\begin{eqnarray}\label{phase}
{\bm F}\to\exp(\rmi\varphi){\bm F}.
\end{eqnarray}
The duality rotation changes the relative proportions of the electric and magnetic field. The presence of electric charges in the real world breaks the symmetry under the duality rotations and introduces the physical distinction between electricity and magnetism. In other words, in the duality plane where the dual rotations (\ref{dual}) take place there is a distinguished direction introduced by the electric charge.

There is some analogy between the change of phase of the RS vector and the change of phase of the wave function in wave mechanics. Namely, the most important physical quantities: the field energy (Hamiltonian) $H$, the momentum ${\bm P}$, and the angular momentum ${\bm M}$ are invariant under the change (\ref{phase}) of the phase of ${\bm F}$. Of course, the phase of ${\bm F}$, unlike the phase of the wave function, is observable once we distinguish between the electric and magnetic field.

The quantities $H$, ${\bm P}$, and ${\bm M}$ are the generators of translations in time and space and rotations. The full list of generators includes also the generator of Lorentz transformation ${\bm N}$. All the generators are built from the energy density,
\begin{eqnarray}\label{end}
\frac{{\bm D}^2}{2\epsilon}+\frac{{\bm B}^2}{2\mu}={\bm F}^*\!\cdot\!{\bm F},
\end{eqnarray}
and the momentum density (Poynting vector divided by $c^2$),
\begin{eqnarray}\label{md}
({\bm D}\times{\bm B})=\frac{1}{\rmi c}({\bm F}^*\times{\bm F}),
\end{eqnarray}
as the following space integrals:
\numparts
\begin{eqnarray}
H=\int\!d^3r\,{\bm F}^*\!\cdot\!{\bm F},\label{gena}\\
{\bm P}=\frac{1}{\rmi c}\int\!d^3r\,({\bm F}^*\times{\bm F}),\label{genb}\\
{\bm M}=\frac{1}{\rmi c}\int\!d^3r\,{\bm r}\times({\bm F}^*\times{\bm F}),\label{genc}\\
{\bm N}=\int\!d^3r\,{\bm r}{\bm F}^*\!\cdot\!{\bm F}\label{gend}.
\end{eqnarray}
\endnumparts
The generator of Lorentz transformation ${\bm N}$ has the simple form of the first moment of energy only at $t=0$. Since it is not a constant of motion, at other times it develops a time-dependent part $-c^2t{\bm P}$.

The square of the RS vector is a complex combination of the two invariants $S$ and $P$ of the electromagnetic field:
\begin{eqnarray}\label{inv}
{\bm F}^2=S+\rmi P=\epsilon\left[\textstyle\frac{1}{2}\left({\bm E}^2-c^2{\bm B}^2\right)+\rmi c{\bm E}\cdot{\bm B}\right].
\end{eqnarray}
Therefore, it is not surprising that the transformation properties of the electromagnetic field expressed in terms of the RS vector have the form of rotations preserving the square of ${\bm F}$. Under ordinary rotations the RS vector transforms as a normal vector:
\begin{eqnarray}\label{rot}
{\bm F}'={\bm F}\cos\varphi+{\bm n}\times{\bm F}\sin\varphi
+{\bm n}\left({\bm n}\cdot{\bm F}\right)(1-\cos\varphi),
\end{eqnarray}
where $\varphi$ is the rotation angle and $\bm n$ is the unit vector of the rotation axis. A pure Lorentz transformation is obtained from this formula by substituting for the rotation angle $\varphi$ the angle of an ``imaginary'' rotation $-\rmi\psi$:
\begin{eqnarray}\label{lor}
{\bm F}'={\bm F}\cosh\psi-\rmi{\bm n}\times{\bm F}\sinh\psi
+{\bm n}\left({\bm n}\cdot{\bm F}\right)(1-\cosh\psi),
\end{eqnarray}
where $\psi=\rm{arctanh}(v/c)$ and $\bm n$ is the unit vector in the direction of the velocity.
\vspace{1.0cm}

\centerline{\Large{\bf Classical theory}}
\addcontentsline{toc}{section}{\Large Classical theory\hspace{20cm}}
\vspace{0.3cm}

\section{Solution of Maxwell equations by Fourier transformation}

The RS vector is ideally suited to use the method of Fourier transformations. Since the RS vector is complex-valued one does not have to impose extra reality conditions on its Fourier transform ${\bm F}({\bm k},t)$, as in the case with electric and magnetic fields. To simplify notation we use the same symbol, but a different argument, to denote the Fourier transform. Taking the three-dimensional Fourier transform of both sides of Maxwell equations one arrives at a simple ordinary differential equation and the transversality condition:
\begin{equation}\label{maxf}
\eqalign{\partial_t{\bm F}({\bm k},t)&=c\,{\bm k}\times{\bm F}({\bm k},t),\\
{\bm k}\cdot{\bm F}({\bm k},t)&=0.}
\end{equation}
The transversality condition requires that the vector ${\bm F}({\bm k},t)$ lies in the subspace orthogonal to ${\bm k}$. We may build the basis in this subspace from two orthogonal real polarization vectors, but it will be more convenient to choose a complex basis ${\bm e}_\pm({\bm k})$ in this subspace built from the eigenvectors of the cross-product,
\begin{eqnarray}\label{cp}
{\bm k}\times{\bm e}_\pm({\bm k}) = \mp\rmi k{\bm e}_\pm({\bm k}),
\end{eqnarray}
where $k=|{\bm k}|$. All normalized vectors that satisfy these equations differ only by phase factors. We choose the phases of these vectors such that ${\bm e}_+({\bm k})={\bm e}({\bm k})$ and ${\bm e}_-({\bm k})={\bm e}^*({\bm k})={\bm e}(-{\bm k})$. The ${\bm e}({\bm k})$ vector will be chosen in the form:
\begin{eqnarray}\label{polarc}
{\bm e}({\bm k})=\frac{{\bm k}\times({\bm m}\times{\bm k})-ik({\bm m}\times{\bm k})}{\sqrt{2}\,k|{\bm m}\times{\bm k}|},
\end{eqnarray}
where $\bm m$ is an arbitrary unit vector. Choosing the $\bm m$ vector along the $z$ axis, one obtains:
\begin{eqnarray}\label{pol}
{\bm e}({\bm k}) = \frac{1}{\sqrt{2}k\sqrt{k_x^2+k_y^2}}\!\left[
\begin{array}{c}
-k_x k_z+\rmi k k_y\\
-k_y k_z-\rmi k k_x\\
k_x^2+k_y^2
\end{array}
\right].
\end{eqnarray}
The polarization vector ${\bm e}({\bm k})$ is dimensionless and is normalized to one, ${\bm e}^*({\bm k})\cdot{\bm e}({\bm k})=1$. In spherical coordinates $k_x=k\cos\varphi\sin\theta,\,k_y=k\sin\varphi\sin\theta,\,k_z=k\cos\theta$, the polari\-zation vector has the form:
\begin{eqnarray}\label{pol1}
{\bm e}(\theta,\varphi)=\frac{1}{\sqrt{2}}\left[
\begin{array}{c}
-\cos\varphi\cos\theta+\rmi \sin\varphi\\
-\sin\varphi\cos\theta-\rmi \cos\varphi\\
\sin\theta
\end{array}
\right].
\end{eqnarray}
Of course, this choice of ${\bm e}({\bm k})$ is not unique because it can always be multiplied by an arbitrary ${\bm k}$-dependent phase factor.

After decomposing ${\bm F}({\bm k},t)$ in the basis of polarization vectors, we solve the resulting ordinary differential equations and we arrive at:
\begin{eqnarray}\label{dec}
{\bm F}({\bm k},t)={\bm e}({\bm k})f_+({\bm k})\rme^{-\rmi\omega t}+{\bm e}(-{\bm k})f_-^*({-\bm k})\rme^{\rmi\omega t},
\end{eqnarray}
where $\omega=ck$. The use of complex conjugation and the sign reversal in the second term may seem awkward, but this notation will be justified later in quantum theory. The inversion of the Fourier transformation gives the following integral formula of the RS vector representing the general solution of the Maxwell equations:
\begin{eqnarray}\label{irep}
{\bm F}(\bm r,t)=\int\!\frac{d^3k}{(2\pi)^{3/2}}{\bm e}(\bm k)\left[f_+(\bm k)\rme^{\rmi\bm k\cdot\bm r-\rmi\omega t}+f_-^*(\bm k)\rme^{-\rmi\bm k\cdot\bm r+\rmi\omega t}\right].
\end{eqnarray}
In the second term we changed the integration variables from $\bm k$ to $-\bm k$. This integral representation is an expansion of the RS vector into circularly polarized monochromatic plane waves,
\begin{eqnarray}\label{pw3}
{\bm F}_{\lambda\bm k}({\bm r},t)={\bm e}({\bm k})\rme^{\rmi\lambda({\bm k}\cdot{\bm r}-\omega t)},
\end{eqnarray}
where $\lambda=\pm 1$ specifies two circular polarization states. The appearance of only {\em one} polarization vector (and not two) in the Fourier representation (\ref{irep}) is a characteristic feature of the RS vector \cite{reply} not shared by the real electromagnetic field vectors $\bm D$ and $\bm B$. This property simplifies many calculations.

Two complex amplitudes $f_\pm$ represent the independent degrees of freedom of the electromagnetic field. They contain all information about the field. In particular, the dynamical quantities (\ref{gena}--d) have the following representation in terms of the functions $f_\lambda({\bm k})$:
\numparts
\begin{eqnarray}
H=\sum_\lambda\int\!\frac{d^3k}{k}\,f_\lambda^*({\bm k})\,\omega\,f_\lambda({\bm k}),\label{genca}\\
{\bm P}=\sum_\lambda\int\!\frac{d^3k}{k}\,f_\lambda^*({\bm k})\,{\bm k}\,f_\lambda({\bm k}),\label{gencb}\\
{\bm M}=\sum_\lambda\int\!\frac{d^3k}{k}\,f_\lambda^*({\bm k})\,\left(\rmi\bm D_\lambda\times{\bm k}+\lambda{\bm n}\right)f_{\lambda}(\bm k),\label{gencc}\\
{\bm N}=\sum_\lambda\int\!\frac{d^3k}{k}\,f_\lambda^*({\bm k})\,\omega\,i\bm D_\lambda\,f_\lambda(\bm k),\label{gencd}
\end{eqnarray}
\endnumparts
where $\bm n$ is the unit vector in the $\bm k$ direction. The symbol ${\bm D}_\lambda$ denotes the covariant derivative on the light cone \cite{qed,bb9},
\begin{eqnarray}
{\bm D}_\lambda={\bm\partial}-i\lambda{\bm\alpha}({\bm k}),\label{cder}\\
{\bm\alpha}({\bm k})= \sum_i e_i^*(\bm k){\bm\partial}e_i(\bm k)=\frac{({\bm m}\!\cdot\!{\bm k})({\bm m}\times{\bm k})}{k\,|{\bm m}\times{\bm k}|^2},\label{alpha}
\end{eqnarray}
where ${\bm\partial}$ denotes the nabla in the ${\bm k}$ space.

To connect the representation (\ref{pw3}) of a plane wave expressed in terms of the RS vector with the formulas found in standard textbooks \cite{jack} we choose the direction of propagation along the $z$ axis which has already been distinguished in the formula (\ref{pol}) for the polarization vector. The simplest choice of ${\bm e}({\bm k})$ in this case gives the RS vector in the form:
\begin{eqnarray}\label{pw4}
{\bm F}_{\lambda\bm k}({\bm r},t)=\frac{1}{\sqrt{2}}\left[
\begin{array}{c}
1\\
\rmi\\
0
\end{array}
\right]\rme^{\rmi\lambda(kz-\omega t)}.
\end{eqnarray}
Separation of this expression into its real and imaginary parts leads to the standard formulas for the electric and magnetic fields of the circularly polarized monochromatic plane waves (we dropped the overall factor),
\begin{eqnarray}\label{pw5}
\fl\qquad\quad{\bm E}({\bm r},t)=\left[
\begin{array}{c}
\cos(kz-\omega t)\\
-\lambda\sin(kz-\omega t)\\
0
\end{array}
\right],\quad {\bm B}({\bm r},t)=\frac{1}{c}\left[
\begin{array}{c}
\lambda\sin(kz-\omega t)\\
\cos(kz-\omega t)\\
0
\end{array}\right].
\end{eqnarray}
One obtains the left-handed or the right-handed circularly polarized waves propagating in the $z$ direction depending on whether $\lambda$ is positive or negative.

\section{Superpotentials}\label{super}

It has been known for more than one hundred years that the use of potentials simplifies the process of solving the Maxwell equations. A systematic method to find and classify all solutions of the Maxwell equations with the use of the RS vector can be based on just {\em one complex scalar function}.

\subsection{Hertz potentials}

We introduce first a complex vector superpotential (complex Hertz potential) ${\bm\Pi}$. The real and imaginary parts of ${\bm\Pi}$ can be identified with the electric and magnetic Hertz potentials ${\bm\Pi}_e$ and ${\bm\Pi}_m$ \cite{hertz,nis}. The RS vector is constructed from ${\bm\Pi}$ according to the formula:
\begin{eqnarray}\label{hertz}
{\bm F}({\bm r},t) = \left[\frac{\rmi}{c}\partial_t+{\bm\nabla}\times\right]({\bm\nabla}\times{\bm\Pi}({\bm r},t)).
\end{eqnarray}
The Maxwell equations obeyed by ${\bm F}$ impose a condition on ${\bm\Pi}$. In order to obtain this condition we act with the differential operator $\left[1/c\,\partial_t+\rmi{\bm\nabla}\times\right]$ on both sides of (\ref{hertz}). The left hand side vanishes on account of (\ref{maxa}) and we obtain the d'Alembert wave equation for the curl of ${\bm\Pi}$,
\begin{eqnarray}\label{wave}
\left(\frac{1}{c^2}\,\partial_t^2-\Delta\right)\left({\bm\nabla}\times{\bm\Pi}({\bm r},t)\right)=0.
\end{eqnarray}
The Hertz potentials deserve the name ``superpotentials'' because the electric and magnetic field vectors are their second derivatives and not the first derivatives as in the case of standard potentials.

There is a lot of freedom in choosing the complex vector ${\bm\Pi}$. The class of gauge transformations of ${\bm\Pi}$ that leave the RS vector unchanged is very large. We shall reduce this freedom by requiring that the vector ${\bm\Pi}$ is derived from a single complex function satisfying the d'Alembert equation. The simplest choice is to assume that ${\bm\Pi}$ has some {\em fixed direction}, ${\bm\Pi}={\bm m}\Phi$. Then,
\begin{eqnarray}\label{whit0}
{\bm F}({\bm r},t)=\left[\frac{\rmi}{c}\partial_t\left({\bm\nabla}\times{\bm m}\right)+{\bm\nabla}\times\left({\bm\nabla}\times{\bm m}\right)\right]\Phi({\bm r},t).
\end{eqnarray}
Choosing the direction of ${\bm m}$ along the $z$ axis, as in the formula (\ref{pol}), one obtains:
\begin{eqnarray}\label{whit}
{\bm F}({\bm r},t)=\left[\begin{array}{c}
\partial_x\partial_z+\rmi/c\,\partial_t\partial_y\\
\partial_y\partial_z-\rmi/c\,\partial_t\partial_x\\
-\partial_x^2-\partial_y^2\
\end{array}
\right]\Phi({\bm r},t).
\end{eqnarray}
For every complex function $\Phi$ satisfying the scalar d'Alembert equation,
\begin{eqnarray}\label{dal}
\left(\frac{1}{c^2}\,\partial_t^2-\Delta\right)\Phi({\bm r},t)=0,
\end{eqnarray}
the RS vector (\ref{whit}) satisfies the Maxwell equations (\ref{maxa}). The formula (\ref{whit}) is a complexified version of the representation introduced a long time ago by Whittaker \cite{whitt,whitt1} so that the real and imaginary parts of $\Phi$ could be called the electric and magnetic Whittaker potentials. The Whittaker construction is particularly useful in the description of beams of electromagnetic radiation since then there is a preferred direction of propagation. The potential $\Phi$ satisfying the d'Alembert equation can be expanded into plane waves,
\begin{eqnarray}\label{irep1}
\fl\qquad\quad{\Phi}(\bm r,t)=\int\!\frac{d^3k}{\sqrt{2}(2\pi)^{3/2}}\frac{1}{k k_\perp}\left[f_+(\bm k)\rme^{\rmi\bm k\cdot\bm r-\rmi\omega t}+f_-^*(\bm k)\rme^{-\rmi\bm k\cdot\bm r+\rmi\omega t}\right].
\end{eqnarray}
Differentiating both sides of this formula according to (\ref{whit}) we reconstruct the Fourier representation (\ref{irep}) of the RS vector. Therefore, every solution can be obtained by the Whittaker construction from an appropriate scalar solution of the d'Alembert equation. The two amplitudes $f_\pm(\bm k)$ in the Fourier expansion of ${\Phi}$ are the amplitudes of circularly polarized parts of the electromagnetic field.

\subsection{Debye potentials}

There is also another choice which is very convenient when dealing with radiation produced by localized sources. In this case one assumes that the superpotential ${\bm\Pi}$ has the {\em radial direction}: ${\bm\Pi}={\bm r}\Phi$. It follows from (\ref{wave}) that $\Phi$ must satisfy the d'Alembert equation (\ref{dal}), as in the Whittaker construction. The real and imaginary parts of $\Phi$ are known as the electric and magnetic Debye potentials \cite{deb}. The RS vector in this case has the form:
\begin{eqnarray}\label{deb}
{\bm F}({\bm r},t)=\left[\frac{1}{c}\partial_t-\rmi{\bm\nabla}\times\right]{\bm L}\Phi({\bm r},t),
\end{eqnarray}
where ${\bm L}$ is the angular momentum operator (up to a factor $\hbar/\rmi$),
\begin{eqnarray}\label{l}
{\bm L}={\bm r}\times{\bm\nabla}.
\end{eqnarray}

The advantages of using scalar functions to represent the electromagnetic field are obvious. Scalar functions represent the {\em independent degrees of freedom}. Moreover, solving the d'Alembert equations for a single function is much simpler than solving several coupled equations. The use of the RS vector results in the replacement of two real scalar functions, electric and magnetic superpotentials, by one complex combination.

\section{Assorted solutions in the new framework}\label{assort}

We shall illustrate now the advantage of using the RS vector by constructing various important solutions of the Maxwell equations with the use of superpotentials constructed from the scalar solutions of the d'Alembert equation. We will be rather careless about the normalization of the RS vector since, due to the linearity of Maxwell equations, the {\em overall} complex amplitude is arbitrary. Obviously, from a given solution $\Phi$ of the d'Alembert equation one may construct two different solutions of Maxwell equations by using either the Whittaker or the Debye construction.

A practical method of finding explicit solutions of the d'Alembert equation is the reduction to ordinary differential equations by the separation of variables. There are 368 coordinate systems in flat space \cite{km} that allow for the separation of variables in the d'Alembert equation. From this vast number of solutions we will select a few examples.

\subsection{Monochromatic waves}

Monochromatic waves are obtained by the separation of the time variable in the d'Alembert equation. Thus, we seek the solution in the form:
\begin{eqnarray}\label{sep}
\Phi({\bm r},t)=\Psi({\bm r})T(t).
\end{eqnarray}
The substitution of this product into the d'Alembert equation results in the following two separate equations:
\numparts
\begin{eqnarray}
\partial_t^2T(t)=-\omega^2T(t),\label{tt}\\
\left(\Delta+\omega^2/c^2\right)\Psi({\bm r})=0\label{helm},
\end{eqnarray}
\endnumparts
where $\omega^2$ is a separation constant. To avoid an unphysical exponential growth of $T(t)$, we assume that $\omega$ is real. The first equation for a given $\omega^2$ has two independent solutions: $\exp(\mp\rmi\omega t)$, hence $\Phi({\bm r},t)$ has two terms:
\begin{eqnarray}\label{genphi}
\Phi({\bm r},t)=\Psi_+({\bm r})\rme^{-\rmi\omega t}+\Psi_-({\bm r})\rme^{\rmi\omega t}.
\end{eqnarray}
The RS vector obtained from such a superpotential (by either Whittaker or Debye construction) has the form:
\begin{eqnarray}\label{2terms}
{\bm F}({\bm r},t)={\bm F}_+({\bm r})\rme^{-\rmi\omega t}+{\bm F}_-({\bm r})\rme^{\rmi\omega t}.
\end{eqnarray}
It follows from Maxwell equations that ${\bm F}_\pm$ are eigenfunctions of the curl operator, known also as the Trkalian fields \cite{trkal},
\begin{eqnarray}\label{trkal}
c{\bm\nabla}\times{\bm F}_\pm({\bm r})=\pm\omega{\bm F}_\pm({\bm r}).
\end{eqnarray}
The real part and the imaginary part are both Trkalian fields.

Monochromatic solutions of Maxwell equations have an interesting property that follows immediately from (\ref{2terms}). The electric and magnetic field vectors at each space point trace ellipses defined by the following parametric representations:
\begin{numparts}
\begin{eqnarray}
{\rm Electric:}\hspace{1cm}\Re({\bm F}_++{\bm F}_-)\cos(\omega t)+\Im({\bm F}_+-{\bm F}_-)\sin(\omega t),\label{pare}\\
{\rm Magnetic:}\hspace{0.8cm}\Im({\bm F}_++{\bm F}_-)\cos(\omega t)-\Re({\bm F}_+-{\bm F}_-)\sin(\omega t).\label{parm}
\end{eqnarray}
\end{numparts}
In general, electric and magnetic ellipses are different, but they coincide when there is only one term present in (\ref{2terms}).

One can further refine the classification of monochromatic waves. One may call a monochromatic solution of Maxwell equations {\em stationary} if its RS vector has the form ${\bm F}({\bm r})\exp(-\rmi\lambda\omega t)$. Each monochromatic wave is a linear combination of two stationary waves as shown in (\ref{2terms}). Stationary waves are direct counterparts of stationary states (energy eigenstates) in quantum mechanics. The energy density and the Poynting vector are time-independent for a stationary wave while in general for a monochromatic wave they oscillate with twice the wave frequency.

The Helmholtz equation (\ref{helm}) for $\Psi({\bm r})$ can be separated in eleven coordinate systems \cite{mf}. Of course, the simplest case is the separation in Cartesian coordinates that leads to the solutions of the d'Alembert equation in the form of monochromatic plane waves (\ref{pw3}).

\subsubsection{Bessel beams.}

Electromagnetic waves described by Bessel functions were known in the context of electromagnetic waves inside the cylindrical wave guides \cite{stratton}. Their use in the description of light beams came later \cite{dme}. Bessel beams are monochromatic and non-diffractive. The mathematical description of Bessel beams is obtained by separation of variables of the Helmholtz equation (\ref{helm}) in cylindrical coordinates. This leads to the following solution of the d'Alembert equation:
\begin{eqnarray}\label{cyl}
\Phi(\rho,\varphi,z,t)=\frac{-\rmi}{kk_\perp}\rme^{\rmi\lambda(k_\parallel z-\omega t+M\varphi)}J_{M}(k_\perp\rho),
\end{eqnarray}
where $k_\parallel$ is the component of $\bm k$ in the direction of propagation,  $k_\perp=\sqrt{k^2-k_\parallel^2}$ and $J_M(k_\perp\rho)$ is the Bessel function. We have chosen the normalization factor in such a way that in the limit, when $k_\perp\to 0$ one obtains monochromatic plane waves (\ref{pw5}). The parameters $k_\parallel, k_\perp$, and $M$ are the analogs of quantum numbers. We shall give a precise formulation of this analogy later when we describe photon states. The meaning of $k_\parallel, k_\perp, M$ and $\lambda$ will then become clear.

The Whittaker construction leads to the following expression for the RS vector of the Bessel beam:
\begin{eqnarray}\label{bess}
\fl\quad\left[\begin{array}{c}
F_x\\
F_y\\
F_z\end{array}\right]
=\rme^{\rmi\lambda(k_\parallel z-\omega t+M\varphi)} \left[\begin{array}{c}\kappa_-\rme^{\rmi\varphi}J_{M+\lambda}(k_\perp\rho)+
\kappa_+\rme^{-\rmi\varphi}J_{M-\lambda}(k_\perp\rho)\\
-\rmi \kappa_-\rme^{\rmi\varphi}J_{M+\lambda}(k_\perp\rho)+
\rmi \kappa_+\rme^{-\rmi\varphi}J_{M-\lambda}(k_\perp\rho)\\
-\rmi \kappa_\perp/k J_{M}(k_\perp\rho)
\end{array}\right],
\end{eqnarray}
where $\kappa_\pm =(1\pm k_\parallel/k)/2$. These formulas slightly simplify when one uses the projections $(F_\rho,F_\varphi,F_z)$ of $F$ on the basis vectors of cylindrical coordinates $(\bm e_r,\bm e_\varphi,\bm e_z)$,
\begin{eqnarray}\label{bess1}
\fl\quad\left[\begin{array}{c}
F_\rho\\
F_\varphi\\
F_z\end{array}\right]
=\rme^{\rmi\lambda(k_\parallel z-\omega t+M\varphi)} \left[\begin{array}{c}\kappa_-J_{M+\lambda}(k_\perp\rho)+
\kappa_+J_{M-\lambda}(k_\perp\rho)\\
-\rmi \kappa_-J_{M+\lambda}(k_\perp\rho)+
\rmi \kappa_+J_{M-\lambda}(k_\perp\rho)\\
-\rmi k_\perp/k J_{M}(k_\perp\rho)
\end{array}\right].
\end{eqnarray}

In the limit $k_\perp\to 0$, only the Bessel function $J_0$ survives. Choosing $M=1,\,\lambda=1$ or $M=-1,\,\lambda=-1$ one obtains circularly polarized plane waves (\ref{pw4}) propagating in the $z$ direction. Bessel beams cannot be exactly realized in Nature since, like plain waves, they carry infinite energy flux---Bessel functions fall off too slowly when $\rho\to\infty$ to make the integral of the Poynting vector over the plane perpendicular to the beam axis finite. However, beams of light resembling Bessel beams in the neighborhood of the propagation axis are now commonly produced in optical laboratories \cite{exper}.

\subsubsection{Multipole fields.}

Multipole fields are very useful in the studies of electro\-magnetic radiation by localized sources. The derivation of the solutions of Maxwell equations in the form of multipole fields is usually a cumbersome task (cf., for example, \cite{jack}). The use of the RS vector expressed in terms of the complex Debye potential significantly simplifies the calculations. All one has to do is to choose the solution of the d'Alembert equation $\Phi$ in the form:
\begin{eqnarray}\label{mult}
\Phi(r,\theta,\varphi,t)=\rme^{-\rmi\lambda\omega t}Y_{lm}(\theta,\varphi)j_l(kr),
\end{eqnarray}
where $Y_{lm}$ is the spherical harmonic and $j_l(kr)$ is the spherical Bessel function and then to perform the differentiations required in evaluating $F$. Treating the expression (\ref{mult}) as  the complex Debye potential in the formula (\ref{deb}) we obtain:
\begin{eqnarray}\label{deb1}
{\bm F}({\bm r},t) = \rme^{-\rmi\lambda\omega t}\left(\lambda k+{\bm\nabla}\times\right)
\left[\begin{array}{c}L_x\\L_y\\L_z
\end{array}\right]Y_{lm}(\theta,\varphi)j_l(kr).
\end{eqnarray}
These are the multipole solutions of Maxwell equations. In the description in terms of the RS vector the existence of two types of multipoles: the electric type and the magnetic type finds its counterpart in the two values of $\lambda$. The formula (\ref{deb1}) has a simple form when expressed in terms of the projections $(F_r,F_\theta,F_\varphi)$ of $\bm F$ on the spherical coordinate versors $(\bm e_r,\bm e_\theta,\bm e_\varphi)$,
\begin{eqnarray}\label{mult1}
\left[\begin{array}{c}
F_r\\
F_\theta\\
F_\varphi
\end{array}\right]
=\frac{1}{r}\left[\begin{array}{c}-l(l+1)\\
-\lambda kr/\sin\theta\,\partial_\varphi-\partial_\theta\partial_r r\\
\lambda kr\partial_\theta-1/\sin\theta\,\partial_\varphi\partial_r r
\end{array}\right]\Phi(r,\theta,\varphi,t).
\end{eqnarray}

The function (\ref{mult}) is well known from wave mechanics where it represents the solution of the time-independent Schr\"odinger equation in free space with quantum numbers $k$, $l$, and $m$. We have chosen the standing waves described by the spherical Bessel functions, but one may also replace them by spherical Hankel functions $h^{(1,2)}_l(kr)$ to describe the outgoing or incoming radiation.

\subsection{Nonmonochromatic beams}

Monochromatic solutions of Maxwell equations are a very convenient theoretical tool, but they are unrealistic since they describe a never ending wave train. In many cases the idealization involved here is harmless, but quite often nonmonochromatic solutions of Maxwell equations are more appropriate. The time-dependence of such solutions is nontrivial and the RS vector comes in handy.

\subsubsection{General plane waves.}

The simplest example of a nonmonochromatic wave is a general plane wave. Such a wave ${\bm F}({\hat{\bm k}}\cdot{\bm r}/c-t)$ propagating in a given direction ${\hat{\bm k}}={\bm k}/k$ is obtained from the general solution (\ref{irep}) by leaving out the angular integration and retaining only the integration over $\omega=kc$,
\begin{eqnarray}\label{spw1}
\fl\qquad{\bm F}({\hat{\bm k}}\cdot{\bm r}/c-t)={\bm e}({\bm k})\int_{0}^{\infty}d\omega\left[ A_+(\omega)\rme^{\rmi\omega({\hat{\bm k}}\cdot{\bm r}/c-t)}+A_-(\omega)\rme^{-\rmi\omega({\hat{\bm k}}\cdot{\bm r}/c-t)}\right],
\end{eqnarray}
where, as usual, the two terms correspond to two circular polarizations. We placed the polarization vector ${\bm e}({\bm k})$ outside the integral because it does not depend on $k$. By allowing $\omega$ to take on positive and negative values, we may write the result as a single integral,
\begin{eqnarray}\label{spw2}
\fl\qquad{\bm F}({\hat{\bm k}}\cdot{\bm r}/c-t)={\bm e}({\bm k})\int_{-\infty}^{\infty}d\omega A(\omega)\rme^{\rmi\omega({\hat{\bm k}}\cdot{\bm r}/c-t)}={\bm e}({\bm k}){\tilde A}({\hat{\bm k}}\cdot{\bm r}/c-t),
\end{eqnarray}
where the wave amplitude ${\tilde A}$ is the Fourier transform of the spectral function $A(\omega)$.

\subsubsection{Gaussian beams.}

Gaussian beams, unlike Bessel beams, are neither mono\-chromatic nor non-diffractive, but they are more realistic; they carry finite energy flux. Gaussian beams are used to describe light emitted by lasers. Very often they are treated as monochromatic, but this is only possible in the paraxial approximation. Here we consider Gaussian beams as exact solutions of Maxwell equations. The RS vector is ideally suited to describe these solutions.

The solutions of Maxwell equations that are not monochromatic form a more diverse collection because one is then not restricted to a rather rigid framework imposed by the Helmholtz equation (\ref{helm}). Gaussian beams can be obtained after the change of variables, $\tau=z+ct,\;\zeta=z-ct$, in the d'Alembert equation. This leads to:
\begin{eqnarray}\label{waw}
\left(4\partial_\tau\partial_\zeta+\Delta_\perp\right)
\Phi({\bm r}_\perp,\tau,\zeta)=0.
\end{eqnarray}
where ${\bm r}_\perp=(x,y)$ and $\Delta_\perp$ is the Laplacian in two-dimensions.

We will seek the solutions of (\ref{waw}) by separating the dependence on $\zeta$,
\numparts
\begin{eqnarray}
\rmi\partial_\zeta\phi(\zeta)=-\mu\phi(\zeta),\label{sep1}\\
\rmi\partial_\tau\psi({\bm r}_\perp,\tau)
=-\frac{1}{4\mu}\Delta_\perp\psi({\bm r}_\perp,\tau).\label{sep2}
\end{eqnarray}
\endnumparts
We have chosen a real value of the separation constant $\mu$ (it can be positive or negative) because any imaginary addition will cause the unphysical exponential growth of the field. The resulting Schr\"odinger-like equation in two dimensions (\ref{sep2}) can be solved by the separation of variables in 26 coordinate systems \cite{bkm}. Gaussian solutions, however, can be obtained without resorting to variable separation. To this end we express the solution of (\ref{sep2}) as a Fourier transform and obtain the following formula for $\Phi$:
\begin{eqnarray}\label{four}
\fl\qquad\quad\Phi({\bm r}_\perp,z,t)=e^{\rmi\mu(z-ct)}\int\!d^2k\,f({\bm k}_\perp)\rme^{\rmi{\bm k}_\perp\cdot{\bm r}_\perp}
\exp\left(-\rmi\frac{{\bm k}_\perp^2}{4\mu}(z+ct)\right),
\end{eqnarray}
where $f({\bm k}_\perp)$ is an arbitrary function. To obtain a simple Gaussian beam we choose $f$ as a pure exponential without polynomial prefactors (such polynomials can be generated later by taking derivatives with respect to ${\bm r}_\perp$),
\begin{eqnarray}\label{gauss}
f({\bm k}_\perp)=\rme^{-{\bm k}_\perp\cdot{\hat K}\cdot{\bm k}_\perp}.
\end{eqnarray}
The complex $2\times2$ matrix ${\hat K}$ is arbitrary but its real part should be positive to produce a convergent integral. With this choice of $f({\bm k}_\perp)$ the Fourier integral can be evaluated and the function $\Phi({\bm r}_\perp,z,t)$ takes on the form (dropping an overall numerical factor):
\begin{eqnarray}\label{gauss1}
\fl\Phi({\bm r}_\perp,z,t)=\frac{\rme^{\rmi\mu(z-ct)}}{\sqrt{{\rm Det}\left\{4{\hat K}+\rmi(z+ct)/\mu{\hat I}\right\}}}
\exp\left(-{\bm r}_\perp\cdot\frac{1}{4{\hat K}+\rmi(z+ct)/\mu{\hat I}}\cdot{\bm r}_\perp\right).
\end{eqnarray}
The simplest Gaussian beam is obtained when ${\hat K}$ is a multiple of the unit matrix, ${\hat K}=(l^2/4){\hat I}$. Then,
\begin{eqnarray}\label{gauss2}
\Phi({\bm r}_\perp,z,t)=\frac{\rme^{\rmi\mu(z-ct)}}{h(z+ct)}
\exp\left(-\frac{{\bm r}_\perp^2}{h(z+ct)}\right),
\end{eqnarray}
where $h(z+ct)=l^2+\rmi(z+ct)/\mu$. The RS vector constructed from this function is:
\begin{eqnarray}\label{gauss3}
\fl{\bm F}(\rho,\varphi,\zeta,\tau)=2\frac{\exp\left(\rmi\mu\zeta-\frac{\rho^2}{h(\tau)}\right)}{h^4(\tau)}
\left[\!\begin{array}{c}
-\rmi\rho \rme^{-\rmi\varphi}\mu h^2(\tau)+\rmi\rho \rme^{\rmi\varphi}(2h(\tau)-\rho^2)/\mu\\
\rho \rme^{-\rmi\varphi}\mu h^2(\tau)+\rho \rme^{\rmi\varphi}(2h(\tau)-\rho^2)/\mu\\
2h(\tau)(h(\tau)-\rho^2)
\end{array}\!\right]\!.
\end{eqnarray}
This compact form shows the efficiency of the description in terms of the RS vector. The formulas for the electric and magnetic field would be much more complicated. However, in order to calculate the energy density and the Poynting vector one does not need the real and imaginary parts; one may use directly the formulas (\ref{end}) and (\ref{md}). We will present here the results for the energy per unit length along the $z$ axis and the energy flux through the $xy$ plane. Keeping the normalization the same as in (\ref{gauss3}) one obtains:
\begin{eqnarray}\label{enpoy}
\fl\qquad\int\!dx dy\,{\bm F}^\dagger\!\cdot\!{\bm F}=\pi\frac{2l^4\mu^4+4l^2\mu^2+3}{l^8\mu^2},\quad
\frac{1}{\rmi}\int\!dx dy\,{\bm F}^\dagger\times{\bm F}=\pi\frac{2l^4\mu^4-3}{l^8\mu^2}.
\end{eqnarray}
These values do not depend on $z$ or $t$ as is required by the continuity equation for the energy and the energy flux. Standard Gaussian beams obtained in the paraxial approximation do not have this property. Note that the beam does not transmit any energy when $l\mu=(3/2)^{1/4}$.

\subsection{Construction of electromagnetic beams as superpositions of simpler solutions}

Monochromatic plane waves, Bessel beams, Gaussian beams, multipole fields are examples of complete sets of solutions: every solution of Maxwell equations can be constructed as a superposition of solutions from a given set. Construction of solutions in this way is most easily done with the use of the RS vector because the building blocks are most often complex functions and postponing the separation of the electric and magnetic parts until the very end greatly simplifies the calculations.

Monochromatic plane waves are the most natural building blocks owing to their very simple mathematical form and their close relation to the Fourier transform. Every solution of Maxwell equations can be built from plane waves according to the formula (\ref{irep}). An expansion into monochromatic plane waves is the best source of information about the beam structure that helps in their construction. It also gives the information about the beam spectrum. Usually one cannot evaluate the integrals over $\bm k$ in a closed form. We shall present here four examples where the integrals over $\bm k$ can be explicitly evaluated. Examples of solutions obtained as superpositions of solutions other than plane waves will also be given.

\subsubsection{Construction of Bessel beams from plane waves.}

Bessel beams are monochro\-matic and the component of the wave vector $k_\parallel$ along the beam axis is fixed. Therefore, there is no integration over $k$ or $k_\parallel$. This means that $k_\perp$ is also fixed. The only remaining integration variable in (\ref{irep}) is the angle $\phi$. The simplest way to construct Bessel beams from plane waves is to start from the following integral representation of the Bessel function:
\begin{eqnarray}\label{besint}
\rmi^{M}\rme^{\rmi M\varphi}J_M(k_\perp\rho)=
\frac{1}{2\pi}\int_0^{2\pi}\!d\phi\,\rme^{\rmi M\phi}\rme^{\rmi k_\perp\rho\cos(\phi-\varphi)}.
\end{eqnarray}
This formula can be extended to the three-dimensional integral by inserting two $\delta$-functions,
\begin{eqnarray}\label{besint1}
\fl\rme^{-\rmi ckt}\rme^{\rmi M\varphi}\rme^{\rmi k_\parallel z}J_M(k_\perp\rho)&=\frac{1}{2\pi\rmi^M k_\perp}\int_0^{\infty}\!\!\!
dq_\perp q_\perp\delta(q_\perp-k_\perp)
\int_{-\infty}^{\infty}\!\!\!dq_\parallel
\delta(q_\parallel-k_\parallel)\nonumber\\
&\times\int_0^{2\pi}\!\!\!d\phi\,
\rme^{\rmi M\phi}\rme^{-\rmi cqt}\rme^{\rmi\left(q_\parallel z+ q_\perp\rho\cos(\phi-\varphi)\right)}.
\end{eqnarray}
The last exponential factor in this formula is nothing else but $\exp(\rmi\bm q\cdot\bm r)$ written in cylindrical coordinates where $\phi$ is the polar angle in momentum space while $\varphi$ is the polar angle in coordinate space. Applying the Whittaker prescription to both sides of this equation we obtain on the left-hand side (up to the factor $-\rmi/kk_\perp$) the Bessel beam (\ref{bess}). On the right-hand side we have a superposition of plane waves whose wave vectors have fixed length and lie on the surface of a cone. The phases of these plane waves are controlled by the factor $\exp(\rmi M\phi)$.

\subsubsection{Construction of Gaussian beams from plane waves.}

Since Gaussian beams are nonmonochromatic, their expansion into plane waves will reveal their spectral composition. To obtain the spectrum it is sufficient to write down the function $\Phi$ given by (\ref{four}) as a three dimensional Fourier transform. This is done by an insertion of an integration over $k_\parallel$ and an appropriate $\delta$ function,
\begin{eqnarray}\label{four1}
\fl\qquad\quad\Phi({\bm r}_\perp,z,t)=\int\!d^3k\,f({\bm k}_\perp)\,\delta\left(k_\parallel+\frac{\bm k_\perp^2}{4\mu}-\mu\right)
\rme^{\rmi({\bm k}\cdot{\bm r}-{\rm sgn}(\mu)\omega t)},
\end{eqnarray}
where we used the equality:
\begin{eqnarray}\label{omega}
\omega/c=\sqrt{k_\perp^2+k_\parallel^2}=\left(|\mu|+\frac{\bm k_\perp^2}{4|\mu|}\right),
\end{eqnarray}
which is valid in the subspace defined by the $\delta$ function in (\ref{four1}). This formula for $\omega$ determines the spectrum of the wave. Note, that for a fixed value of $\mu$ the frequencies $\omega$ lower than $c|\mu|$ are excluded. This property is valid not only for Gaussian beams, but for all beams obtained from the solutions of the Schr\"odinger equation (\ref{sep2}).

\subsubsection{Construction of exponential beams from Bessel functions.}

In the previous two examples we constructed solutions of the d'Alembert equation from plane waves. One may also construct a new solution as a superposition of other solutions, for example Bessel solutions. This possibility is exemplified by exponential beams \cite{bb5}. Exponential beams are similar to Gaussian beams, but they decrease more slowly with the distance from the axis. They are obtained as a superposition of functions (\ref{cyl}) with fixed values of $M$ and $k_\parallel$ but with the varying value of $k_\perp$. Without loss of generality one may assume that $\lambda=1$ since the change of the sign of $\lambda$ results only in complex conjugation. Choosing the dimensionless integration variable as $w=\sqrt{1+(k_\perp/k_\parallel)^2}$ one obtains:
\begin{eqnarray}\label{genbeam}
\fl\qquad\chi_{Mk_\parallel}(\rho,\varphi,z,t)
= \rme^{\rmi\left(M\varphi+k_\parallel z\right)}\int_1^\infty\!dw\,g(w)
\rme^{-\rmi c|k_\parallel|wt}J_M(|k_\parallel|\rho\sqrt{w^2-1}).
\end{eqnarray}
where $g(w)$ is an arbitrary weight function. There are several choices of this function for which the integral can be analytically evaluated \cite{bb5}. For beam-like solutions, the longitudinal component $k_\parallel$ should not be spread out too much and also it must be much larger than the transverse component $k_\perp$. The integration over $w$ in (\ref{genbeam}) may be performed only in a few cases. The simplest choice is:
\begin{eqnarray}\label{weight}
g(w)=(w^2-1)^{M/2}\rme^{-c|k_\parallel|\tau w},
\end{eqnarray}
where $\tau$ is a parameter with the dimension of time whose inverse determines the width of the beam spectrum. The formula 6.645.2 in~\cite{gr} gives:
\begin{eqnarray}\label{expbeam3}
\chi_{Mk_\parallel}(\rho,\varphi,z,t)=\rme^{\rmi(M\varphi+k_\parallel z)}\sqrt{\frac{2}{\pi\vert k_\parallel\vert}}\frac{\rho^M K_{M+1/2}\left(\vert k_\parallel\vert s\right)}{s^{M+1/2}},
\end{eqnarray}
where
\begin{eqnarray}\label{defs}
s=\sqrt{\rho^2 - c^2(t-\rmi\tau)^2}.
\end{eqnarray}
The MacDonald function $K_{M+1/2}$ for half-integer values of the index reduces to an exponential with some prefactors. In the simplest case when $M=0$ one obtains:
\begin{eqnarray}
\chi_{0 k_\parallel}(\rho,\varphi,z,t) = \rme^{\rmi k_\parallel z}
\frac{\rme^{-\vert k_\parallel\vert s}}{\vert k_\parallel\vert s}.
\end{eqnarray}
The RS vector obtained from this function by either Whittaker or Debye construction will exhibit an exponential fall-off with the increasing value of $\rho$.

\subsection{Construction of solutions from an arbitrary complex function}\label{g}

A large class of solutions can be obtained by the Whittaker construction starting from the solution of the d'Alembert equation $\Phi$ in the following form:
\begin{eqnarray}\label{arb}
\Phi(r,t)=\frac{g(ct-r)}{r}-\frac{g(ct+r)}{r},
\end{eqnarray}
where $g$ is an arbitrary complex function. The first term describes an outgoing spherical wave; the second term describes an incoming spherical wave. Both waves are singular at $r=0$ but their difference is regular. Perhaps the simplest solution of this kind is:
\begin{eqnarray}\label{sim}
\Phi(r,t)=\rme^{-\rmi\omega t}\frac{\sin kr}{r}.
\end{eqnarray}
A more elaborate example was studied in connection with the problem of photon localization in \cite{exp} where the function $g(\tau)$ was chosen in the form:
\begin{eqnarray}\label{exploc}
g(\tau)=\exp(-2\sqrt{1+\rmi\tau}).
\end{eqnarray}
In Sec.~\ref{fund} we give another important example of a solution defined by a very simple choice of $g(\tau)$.

\subsection{Generation of new solutions by an imaginary shift}

Maxwell equations are invariant under all Poincar\'e transformations, but this obviously does not lead to any {\em essentially new} solutions. However, as pointed out by Trautman \cite{traut}, all linear relativistic equations are invariant under the complex Poincar\'e transformations and this fact may be used to generate entirely new solutions. We shall show that even in the simplest case of an imaginary translation we obtain a different solution. The use of the RS vector is almost mandatory here since imaginary shift cannot be applied to real field vectors. We give below two examples of an imaginary shift. In the first case we add an imaginary space-like vector and in the second case it will be a time-like vector.

\subsubsection{Complexified Coulomb field.}\label{coul}

\begin{figure}
\centering
\includegraphics[width=9cm,height=5.5cm]{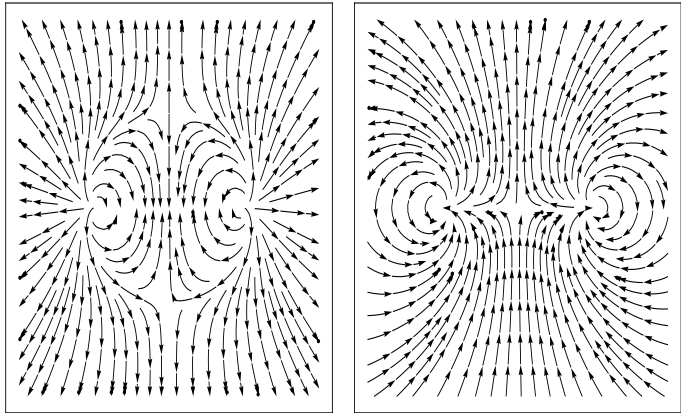}
\caption{Lines of the electric field (left) and the magnetic field (right)\\in the $y=0$ plane drawn for the complexified Coulomb solution. The vector ${\bm m}$ point in the vertical direction.}\label{fig2}
\end{figure}

The simplest, but a highly nontrivial example of a new solution of Maxwell equations is obtained by an imaginary translation ${\bm r}\to {\bm r}+\rmi{\bm m}$ in the Coulomb field \cite{etn}. This produces the RS vector of the form:
\begin{eqnarray}\label{etn}
\fl\qquad\qquad\frac{{\bm r}+\rmi{\bm m}}{({\bm r}^2-{\bm m}^2+2\rmi{\bm m}\!\cdot\!{\bm r})^{3/2}}
=\frac{{\bm r}\cos\psi+{\bm m}\sin\psi}{R^3}+\rmi\frac{{\bm m}\cos\psi-{\bm r}\sin\psi}{R^3},
\end{eqnarray}
where
\begin{eqnarray}\label{rpsi}
\fl\qquad R=\left[({\bm r}^2-{\bm m}^2)^2+4({\bm m}\!\cdot\!{\bm r})^2\right]^{1/4},\quad
\psi=\frac{3}{2}\arg\left({\bm r}^2-{\bm m}^2+2\rmi{\bm m}\!\cdot\!{\bm r}\right).
\end{eqnarray}

The electric and magnetic field lines as seen in figure  \ref{fig2} exhibit quite an intricate form. Since the fields possess cylindrical symmetry, the field configuration in the $y=0$ plane carries full information. Far from the source the electric field approaches the field of a point charge and the magnetic field approaches the field of a magnetic dipole. Since the expression $({\bm r}^2-{\bm m}^2+2\rmi{\bm m}\!\cdot\!{\bm r})^{3/2}$ is double-valued, one must introduce a prescription how this expression is to be evaluated. To keep the fields continuous for $r>|{\bm m}|$ one must introduce discontinuities on the disk of radius $|{\bm m}|$, as shown in figure  \ref{fig2}. The discontinuity of the normal component of the electric field gives the surface charge $2(|{\bm m}|^2-x^2-y^2)^{-3/2}$ and the discontinuity of the tangential component of the magnetic field gives the circular surface current $2(|{\bm m}|^2-x^2-y^2)^{-3/2}$ flowing in the azimuthal direction \cite{gerry}.

\subsubsection{Complexified fundamental solution of the d'Alembert equation.}\label{fund}

A second example of a new solution generated by an imaginary shift is the one studied by Synge \cite{synge}. It can be obtained by the Whittaker construction from the following complexified solution of the d'Alembert equation:
\begin{eqnarray}\label{synge}
\Phi({\bm r},t)=\frac{1}{c^2(t-\rmi a)^2-{\bm r}^2},
\end{eqnarray}
where $a$ is a real constant that defines the scale of this solution. For $a=0$ this is the Hadamard fundamental solution of the d'Alembert equation singular on the light cone, but when $a\neq 0$ this function is analytic in all spacetime. This function is perhaps the simplest solution of the d'Alembert equation with finite energy.

\begin{figure}
\centering
\includegraphics[width=13cm,height=3cm]{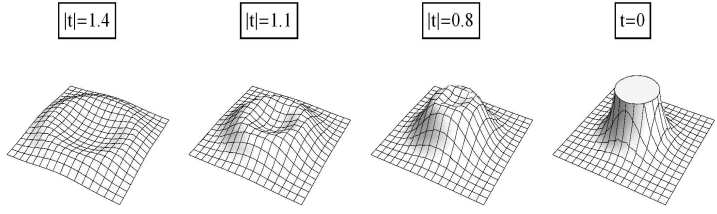}
\caption{Energy density in the $z=0$ plane as a function of time.}\label{fig3}
\end{figure}

The electric and magnetic fields for the Synge solution are fairly complicated, but the energy density and the Poynting vector are given by simple formulas ($c=1$):
\begin{eqnarray}\label{syngenergy}
{\bm F}^*\!\cdot\!{\bm F}
=16\frac{\left(a^2-t^2+x^2+y^2+z^2\right)^2
+8t^2\left(x^2+y^2\right)+4a^2t^2}
{d},\\
\frac{1}{\rmi}{\bm F}^*\times{\bm F}
=\frac{64t}{d}\left[\begin{array}{c}
x\left(a^2+t^2+x^2+y^2-z^2\right)+2azy\\
y\left(a^2+t^2+x^2+y^2-z^2\right)-2azx\\
2z\left(x^2+y^2\right)
\end{array}\right],
\end{eqnarray}
where $d=\left(\left(a^2-t^2+x^2+y^2+z^2\right)^2+4a^2t^2\right)^3$. Note that the energy density is an even function of time while the Poynting vector is an odd function. The behavior of these functions in the $z=0$ plane explains very well the salient properties of the solution. In figure  \ref{fig3} we show the energy density for four values of time that characterizes a collapsing/expanding shell. This figure should be viewed from left to right for negative values of time ending at $t=0$ when the collapse stops (the Poynting vector vanishes). For positive values of time, starting with the rightmost plot, we see the expansion of the wave. The Poynting vector in the $z=0$ plane has no $z$ component and describes the energy flowing towards the center for negative values of time and away from the center for positive values.
The complexified fundamental solution is also directly related to the class of solutions described in Sec.~\ref{g}. The solution of the d'Alembert equation (\ref{synge}) can be written in the form (\ref{arb}) with the function $g(\tau)$ chosen in the form $g(\tau)=\textstyle{\frac{1}{2}}/(\tau-ia)$.

It is worth noting that the solution of Maxwell equations obtained from the function (\ref{synge}) shows up in several unexpected places. In particular it appears in the uncertainty relation for photons \cite{bb8}. Another distinct property of this solution is that it can be obtained from the standard plane wave by applying a conformal transformation \cite{bb3}.

\section{Vortex lines of the electromagnetic field}\label{vortex}

Another area of research where the description of the electromagnetic field in terms of complex functions is necessary is the study of vortex lines: the complex RS vector is indispensable in these studies. Vortex lines are intimately connected with orbital angular momentum and the solutions of wave equations carrying orbital angular momentum are most naturally described by complex functions. The signature of a nonvanishing orbital angular momentum is the complex factor $\rme^{\rmi M\varphi}$. This factor is singular along a line and the removal of the singularity requires the vanishing of the wave function. The mere vanishing of the wave function, however, is not enough to create a vortex line; there must be also the circulation of velocity.

While there is no ambiguity in nonrelativistic quantum mechanics as to how to define vortex lines, the problem does not have an obvious solution for multi-component wave functions. For the one-component wave functions in nonrelativistic quantum mechanics vortex lines appear generically \cite{bb11}. For the multi-component wave functions, such as the RS vector, the requirement that {\em all} components vanish can be satisfied only on rare occasions. The most notable examples are Bessel beams; all components of the RS vector vanish for $|M|\ge2$ along the $z$ axis. The representation of Bessel beams in cylindrical coordinates (\ref{bess1}) exhibits the factor $\rme^{\rmi M\varphi}$ clearly indicating the presence of the vortex line. Other examples of vortex lines are found in null solutions of Maxwell equations \cite{bb3}. These are the solutions with the vanishing square ${\bm F}^2$ of the RS vector discussed in the next section.
\begin{figure}
\centering
\includegraphics[width=7cm,height=7cm]{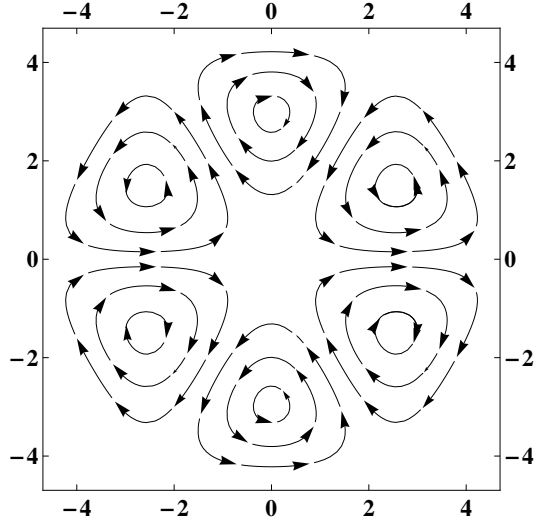}
\caption{Streamlines of the velocity field for the superposition of three\\
circularly polarized plane waves moving in mutually orthogonal directions.\\ This pattern is periodically repeated in vertical and horizontal directions.}\label{fig4}
\end{figure}

The generic vortex lines, i.e. those that are present in almost any solution of Maxwell equations can be obtained by replacing the quantum-mechanical wave function $\psi$ by the square of the RS vector \cite{bb3}. Since ${\bm F}^2$ is one complex function, the equations ${\rm Re}{\bm F}^2=0$ and ${\rm Im}{\bm F}^2=0$ define two surfaces. The intersection of these surfaces defines a line. The velocity field is defined by analogy with the quantum-mechanical case as:
\begin{eqnarray}\label{vel}
u_\mu = \frac{({\bm F}^2)^*\partial_\mu{\bm F}^2-{\bm F}^2\partial_\mu({\bm F}^2)^*}{2\rmi\vert{\bm F}^2\vert^2}=\frac{S\partial_\mu P-P\partial_\mu S}{S^2+P^2}.
\end{eqnarray}
We are able to define a genuine velocity four-vector since ${\bm F}^2$ is a relativistic scalar field (\ref{inv}). As an illustration we calculate the velocity field for the RS vector given by the superposition of three monochromatic circularly polarized plane waves propagating ing in mutually orthogonal directions,
\begin{eqnarray}\label{sup}
{\bm F}=\rme^{-\rmi\omega t}\left[(\bm l+\rmi\bm m)\rme^{\rmi\bm n\cdot\bm k}+(\bm n+\rmi\bm l)\rme^{\rmi\bm m\cdot\bm k}+(\bm m+\rmi\bm n)\rme^{\rmi\bm l\cdot\bm k}\right].
\end{eqnarray}
The three vectors $\bm l,\,\bm m$ and $\bm n$ form an orthonormal set and this results in a very simple form of $\bm F^2$ and subsequently of $u_\mu$, described in detail in \cite{bb2}. Vortex lines are aligned with the vector $\bm l+\bm m+\bm n$. In figure  \ref{fig4} we show the streamlines of the velocity field in the plane orthogonal to this vector $\bm l+\bm m+\bm n$. Vortex lines form a regular stationary lattice with the same number of vortices with clockwise and with anti-clockwise circulation.

\section{Knotted fields}\label{null}

\begin{figure}
\centering
\includegraphics[width=11cm,height=7cm]{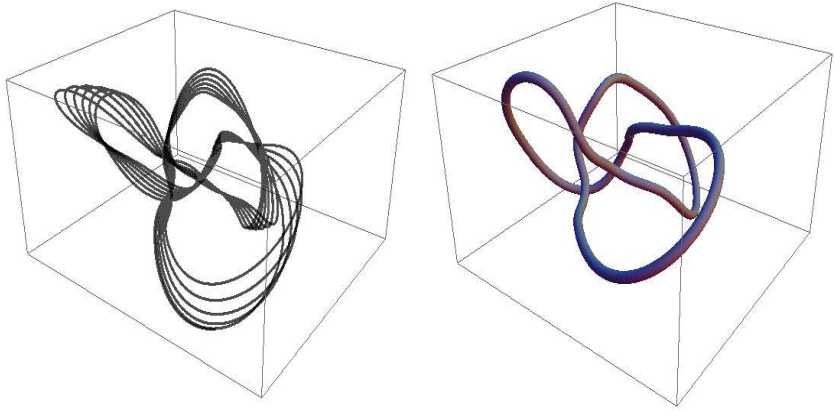}
\caption{Lines of the electric field corresponding to the choice $p=2$ and $q=3$. The plot on the right shows a single trefoil artificially enhanced for clarity. In the plot on the left the line of the electric field was continued much further showing a rolling trefoil (the field line does not exactly close).}\label{fig5}
\end{figure}
There is a special category of solutions of the Maxwell equations that were even given a special name of {\em pure electromagnetic waves} in the early days of modern electromagnetic theory. Nowadays they are called null waves because both invariants (\ref{inv}) are zero (the square of the RS vector vanishes). The simplest example of a null wave is a general plane wave (\ref{spw2}). Bateman \cite{bateman} gave a neat prescription how to construct a null wave in terms of the RS vector. Take a pair of complex functions $\alpha$ and $\beta$ of spacetime coordinates. The RS vector defined as:
\begin{equation}
\mathbf{F}=\nabla\alpha\times\nabla\beta \label{fbtm}
\end{equation}
will be a null solution of the Maxwell equations if $\alpha$ and $\beta$  satisfy the following conditions:
\begin{eqnarray}
{\bm\nabla}\alpha\times{\bm\nabla}\beta=\rmi\left(\partial_{t}\alpha{\bm\nabla}\beta
-\partial_{t}\beta{\bm\nabla}\alpha\right).\label{ab}
\end{eqnarray}
One may check using (\ref{ab}) that the RS vector (\ref{fbtm}) satisfies the Maxwell equations. It can be shown \cite{hogan} that every null solution of the Maxwell equations can be represented in this form. Bateman construction is particularly useful in the description of electromagnetic waves whose field lines possess interesting topological properties. Let us first note that each pair of functions $\alpha$ and $\beta$ satisfying (\ref{ab}) defines a whole family of null solutions whose members are described by the formula
\begin{equation}
\mathbf{F}=h(\alpha,\beta)\nabla\alpha\times\nabla\beta,\label{family}
\end{equation}
where $h(\alpha,\beta)$ is an arbitrary function. Note that we may represent every such RS vector again in the form (\ref{fbtm}) by replacing the original functions $\alpha$ and $\beta$ by $f(\alpha,\beta)$ and $g(\alpha,\beta)$, where the new functions are so chosen (there are many such choices) that $h=\partial_\alpha f\partial_\beta g-\partial_\alpha g\partial_\beta f$.

\begin{figure}
\centering
\includegraphics[width=6cm,height=6cm]{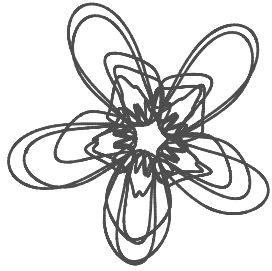}
\caption{Lines of the electric field corresponding to the choice $p=2$ and $q=5$. The plot shows the bundle of electric field viewed from the top (along the $z$-axis).}\label{fig6}
\end{figure}
As an example we shall consider the RS vector built according to (\ref{family}) from the following functions $\alpha$, $\beta$, and $h$ \cite{kbpi}:
\begin{equation}\label{abh}
\fl\alpha=1-2\frac{t-\rmi -z}{x^2 + y^2 + z^2 - (t-\rmi)^2},\; \beta=\frac{2(x-\rmi y)}{x^2 + y^2 + z^2 - (t-\rmi)^2},\; h(\alpha,\beta)=\alpha^{p-1}\beta^{q-1},
\label{example}
\end{equation}
where $p$ and $q$ are natural numbers. In the simplest case, $p=1,\,q=1$, the field lines form collections of linked circles. Topologically more intricate field configurations are obtained when $p$ and $q$ are coprime. For example for $p=2$ and $q=3$ the field lines form a rolling trefoil shown in figure  \ref{fig5}. For $p=2$ and $q=5$ we obtain an even more elaborate bundle of field lines shown in figure  \ref{fig6}. A detailed discussion of these knotted field configurations can be found in \cite{kbpi}. Knotted electromagnetic field configurations not only have an interesting mathematical structure but they were considered as models of ball lighting \cite{ran}.

\section{The RS vector in a general coordinate system}\label{genco}

In this section we show that the usefulness of the RS vector extends to arbitrary coordinate systems (even in curved spacetime). The Maxwell equations in this case can also be reduced to the equations for one complex vector.

The original Maxwell equations (\ref{max0a}--d) hold in any coordinate system also in curved spacetime. The nabla operator that appears in these equations denotes the derivatives with respect to the chosen coordinates and the field vector have components along the coordinates. For example, in spherical coordinates the nabla stands for $(\partial_r,\partial_\theta,\partial_\varphi)$. The whole information about the coordinate system is in the constitutive relations between the $({\bm D},{\bm B})$ pair and the $({\bm E},{\bm H})$ pair. The constitutive relations (cf. for example \cite{pwf,pwf0,jp}) involve the components of the metric tensor $g_{\mu\nu}$ and its inverse $g^{\mu\nu}$ (the Greek indices run from 0 to 3),
\numparts
\begin{eqnarray}
E_i=\frac{1}{\,g^{00}}\left(-g_{ij}/\sqrt{-g}\frac{D^j}{\epsilon}+ cg^{0k}\epsilon_{kij}B^j\right),\label{cr01}\\
H_i=\frac{1}{\,g^{00}}\left(-g_{ij}/\sqrt{-g}\frac{B^j}{\mu}- cg^{0k}\epsilon_{kij}D^j\right),\label{cr02}
\end{eqnarray}
\endnumparts
where we used the summation convention over repeated indices (Latin indices run from 1 to 3) and $g$ is the determinant of the $g_{\mu\nu}$ matrix. The space with a nonvanishing space-time component $g_{0k}$ of the metric tensor is bi-anisotropic because the electric and magnetic components are mutually coupled through the constitutive relations. The vector components in these formulas are the coordinate components not the physical components defined as projections on an orthonormal set of basis vectors that were used so far.

The expressions (\ref{cr01}--b) can also be rewritten in terms of complex vectors $\bm{\mathcal F}$ and $\bm{\mathcal G}$ built according to the formulas (\ref{rs}) and (\ref{grs}), but with the Cartesian vectors $({\bm D},{\bm B})$ and $({\bm E},{\bm H})$ replaced by their counterparts in a general coordinate system,
\numparts
\begin{eqnarray}
{\mathcal G}_i={\mathcal E}_{ij}{\mathcal F}^j
=\frac{1}{\,g^{00}}\left(-g_{ij}/\sqrt{-g}+\rmi g^{0k}\epsilon_{kij}\right){\mathcal F}^j,\label{cr1}\\
{\mathcal F}^i={(\mathcal E^{-1})}^{ij}{\mathcal G}_j
=\frac{1}{\,g_{00}}\left(-g^{ij}\sqrt{-g}-\rmi g_{0k}\epsilon^{kij}\right){\mathcal G}_j.\label{cr2}
\end{eqnarray}
\endnumparts
We use new symbols to denote these vectors because they should not be identified with their Cartesian counterparts. Using Cartesian coordinates we have not distinguished between covariant and contravariant components; this distinction is necessary in curvilinear coordinates.

The easiest way to find the relations (\ref{cr01}--b) is to use the variational principle. The Lagrangian density in spacetime endowed with the metric tensor $g_{\mu\nu}$ has the form
\begin{eqnarray}\label{lag}
{\mathcal L}=-\textstyle{\frac{1}{4}}
\sqrt{-g}\,g^{\mu\lambda}g^{\nu\rho}f_{\mu\nu}f_{\lambda\rho},
\end{eqnarray}
where $f_{\mu\nu}$ is built from the four-vector potential $f_{\mu\nu}=\partial_\mu A_\nu-\partial_\nu A_\mu$. The vectors $({\bm E},{\bm B})$ are obtained from the components of $f_{\mu\nu}$ as follows: $E_k=f_{0k},\,B^k=-f_{ij}$. The form of the Lagrangian (\ref{lag}) is a unique generalization of the expression in the Cartesian coordinates to an arbitrary coordinate system. The Maxwell equations in the tensor form derived from this Lagrangian are:
\numparts
\begin{eqnarray}
\partial_\mu f_{\nu\lambda}+\partial_\lambda f_{\mu\nu}+\partial_\lambda f_{\mu\nu}=0,\label{max1}\\
\partial_\mu h^{\mu\nu}=0,\label{max2}\\
h^{\mu\nu}=-2\frac{\partial{\mathcal L}}{\partial f_{\mu\nu}}=\sqrt{-g}\,g^{\mu\lambda}g^{\nu\rho}f_{\lambda\rho}\,.
\label{max3}
\end{eqnarray}
\endnumparts
From these equations we can read of the relations (\ref{cr01}--b), after the identification: $D^k=h^{k0}$ and $H_k=-h^{ij}$.

The matrix ${\mathcal E}_{ij}$, that determines the constitutive relations (\ref{cr1}--b) appears also in Maxwell equations for the RS vector. These equations may be expressed either in terms of $\bm{\mathcal F}$ or in terms of $\bm{\mathcal G}$,
\numparts
\begin{eqnarray}
\rmi\partial_t{\mathcal F}^i
=\epsilon^{ijk}\partial_j{\mathcal E}_{kl}{\mathcal F}^l,\label{max3a}\\
\rmi\partial_t{({\mathcal E}^{-1})}^{ij}{\mathcal G}_j
=\epsilon^{ijk}\partial_j{\mathcal G}_k.\label{max3b}
\end{eqnarray}
\endnumparts
In each case one has only {\em one complex vector} to describe the electromagnetic field as was announced at the beginning of this section.

The definition of the RS vector in a general coordinate system involves a subtle point that should be elucidated. We shall start by writing the formula for the Hamiltonian in a general coordinate system,
\begin{eqnarray}
H&=\textstyle\frac{1}{2}\int d^3r\left({\bm D}\!\cdot\!{\bm E}+{\bm B}\!\cdot\!{\bm H}\right)=\textstyle\frac{1}{2}\int dx^1dx^2dx^3\left(D^iE_i+B^iH_i\right)\nonumber\label{en0}\\
&=\int dx^1dx^2dx^3\left(-\frac{D^ig_{ij}D^j}{2\epsilon}-\frac{B^ig_{ij}B^j}{2\mu}
+cg^{0k}\epsilon_{kij}D^iB^j\right)\nonumber\label{en1}\\
&=\int dx^1dx^2dx^3 \mathcal F^{*i}{\mathcal E}_{ij}{\mathcal F}^j,\label{en2}
\end{eqnarray}
where we used the constitutive relations (\ref{cr01}--b). The Hamiltonian coincides with the energy when $g_{00}=1$ and all components $g_{0k}$ vanish. These conditions are satisfied in the simple case of flat space that will be considered first. The subtle point is the presence of different volume elements in the expressions (\ref{gena}) and (\ref{en2}). In the first case it is $d^3r$; in the second case it is $dx^1dx^2dx^3$. After the transformation to the coordinates $x^i$ the volume element $d^3r$ contains in addition to $dx^1dx^2dx^3$ also the Jacobian of the transformation. For example, in spherical coordinates we have $d^3r=r^2\sin\theta\,dr d\theta d\varphi$. The Jacobian does not appear explicitly in the expression (\ref{en1}) for the energy, it is contained in the product of the RS vectors.

\subsection{Bessel beams in cylindrical coordinates}

We shall apply now this formulation to the cylindrical coordinate system in flat space. The time variable in this case is not affected and the spatial line element is:
\begin{eqnarray}
ds^2=d\rho^2+\rho^2d\varphi^2+dz^2.\label{ds2}
\end{eqnarray}
Hence, $\sqrt{-g}=\rho$ and the constitutive relations read:
\begin{eqnarray}
\left({\mathcal G}_\rho,{\mathcal G}_\varphi,{\mathcal G}_z\right)=\left(\frac{{\mathcal F}^\rho}{\rho},\rho {\mathcal F}^\varphi,\frac{{\mathcal F}^z}{\rho}\right).\label{cr3}
\end{eqnarray}
The equations for $\bm{\mathcal F}$ become:
\numparts
\begin{eqnarray}
\rmi\frac{1}{c}\partial_t{\mathcal F}^\rho=\partial_\varphi\frac{{{\mathcal F}^z}}{\rho}
-\partial_z\rho {\mathcal F}^\varphi,\label{mcyla}\\
\rmi\frac{1}{c}\partial_t{\mathcal F}^\varphi=\partial_z\frac{{{\mathcal F}^\rho}}{\rho}
-\partial_\rho\frac{{\mathcal F}^z}{\rho},\label{mcylb}\\
\rmi\frac{1}{c}\partial_t{\mathcal F}^z=\partial_\rho\rho{{\mathcal F}^\varphi}
-\partial_\varphi\frac{{\mathcal F}^\rho}{\rho},\label{mcylc}.
\end{eqnarray}
\endnumparts

Our aim now is to reproduce the results for the Bessel beam. To this end we shall seek the solution of equations (\ref{mcyla}--c) in the form:
\begin{eqnarray}
{\bm{\mathcal F}}(\rho,\varphi,z,t)=\rme^{\rmi\lambda(k_\parallel z-\omega t+M\varphi)}{\bm{\mathcal F}}(\rho).\label{form}
\end{eqnarray}
Upon substituting this Ansatz into (\ref{mcyla}--c), we obtain:
\numparts
\begin{eqnarray}
k{\mathcal F}^\rho=\frac{\rmi M}{\rho}{{\mathcal F}^z}
-\rmi k_\parallel\rho{\mathcal F}^\varphi,\label{cyla}\\
k{\mathcal F}^\varphi=\frac{\rmi k_\parallel}{\rho}{{\mathcal F}^\rho}
-\lambda\partial_\rho\frac{1}{\rho}{\mathcal F}^z,\label{cylb}\\
k{\mathcal F}^z=\lambda\partial_\rho\rho{{\mathcal F}^\varphi}
-\frac{\rmi M}{\rho}{\mathcal F}^\rho.\label{cylc}
\end{eqnarray}
\endnumparts
Next, we solve  the first two equations (\ref{cyla}) and (\ref{cylb}) with respect to ${\mathcal F}^\rho$ and ${\mathcal F}^\varphi$,
\numparts
\begin{eqnarray}
{\mathcal F}^\rho=\frac{\rmi}{k_\perp^2}\left(kM+\lambda\rho k_\parallel\partial_\rho\right)\frac{{\mathcal F}^z}{\rho},\label{c1a}\\
\rho{\mathcal F}^\varphi=-\frac{1}{k_\perp^2}\left(k_\parallel M+\lambda k\rho\partial_\rho\right)\frac{{\mathcal F}^z}{\rho}.\label{c1b}
\end{eqnarray}
\endnumparts
After the substitution of these expression into the third equation we obtain the Bessel equation for the function ${\mathcal F}^z/\rho$,
\begin{eqnarray}
\left(\frac{1}{\rho}\partial_\rho\rho\partial_\rho
-\frac{M^2}{\rho^2}+k_\perp^2\right)\frac{{\mathcal F}^z}{\rho}=0.\label{cyl1b}
\end{eqnarray}
Comparing this with the solution for the Bessel beam (\ref{bess1}) one can see that ${\mathcal F}^z$ in cylindrical coordinates differs by a factor of $\rho$ from the Cartesian $F_z$. By comparing the expression for the energy (\ref{gena}) and (\ref{en2}) one can establish the full connection between ${\bf\mathcal F}$ and ${\bf F}$,
\begin{eqnarray}
\left({\mathcal F}^\rho,{\mathcal F}^\varphi,{\mathcal F}^z\right)=\left(\rho F_\rho,F_\varphi,\rho F_z\right).\label{conn}
\end{eqnarray}

\subsection{Multipole fields in spherical coordinates}

The spatial line element in spherical coordinates is:
\begin{eqnarray}
ds^2=dr^2+r^2d\varphi^2+r^2\sin\theta^2d\varphi^2.\label{dss}
\end{eqnarray}
Hence, $\sqrt{-g}=r^2\sin\theta$ and the constitutive relations read:
\begin{eqnarray}
\left({\mathcal G}_r,{\mathcal G}_\theta,{\mathcal G}_\varphi\right)=\left(\frac{{\mathcal F}^r}{r^2\sin\theta},\frac{{\mathcal F}^\theta}{\sin\theta},\sin\theta{\mathcal F}^\varphi\right).\label{crs}
\end{eqnarray}
The equations for $\bm{\mathcal F}$ after the separation of the variables $t$ and $\varphi$ became:
\numparts
\begin{eqnarray}
k{\mathcal F}^r=\partial_\theta\sin\theta{{\mathcal F}^\varphi}
-\frac{\rmi m{\mathcal F}^\theta}{\sin\theta},\label{mspha}\\
k{\mathcal F}^\theta=\frac{\rmi m{{\mathcal F}^r}}{r^2\sin\theta}
-\partial_r\sin\theta{\mathcal F}^\varphi,\label{msphb}\\
k{\mathcal F}^\varphi=\partial_r\frac{{{\mathcal F}^\theta}}{\sin\theta}
-\partial_\theta\frac{{\mathcal F}^r}{r^2\sin\theta}.\label{msphc}
\end{eqnarray}
\endnumparts
These equations are satisfied by the expressions (\ref{mult1}) provided one takes into account the following relations between $\bm{\mathcal F}$ and $\bm F$:
\begin{eqnarray}
\left({\mathcal F}^r,{\mathcal F}^\theta,{\mathcal F}^\varphi\right)=\left(r^2\sin\theta F^r,r\sin\theta F^\theta,rF^\varphi\right),\label{conn1}
\end{eqnarray}
which may be obtained from the two formulas for the energy as in the derivation of (\ref{conn}).

\subsection{Electromagnetic field in curved space}

We consider here, as a simple example, the curved space in the vicinity of a black hole of mass $M$. We use the standard form of the Schwarzschild metric,
\begin{eqnarray}
\fl\qquad c^2d\tau^2=c^2\left(1-r_s/r\right)dt^2
-\frac{dr^2}{1-r_s/r}-r^2\left(d\theta^2+\sin^2\theta\, d\varphi^2\right),\label{sch}
\end{eqnarray}
where $r_s=2GM/c^2$ is the Schwarzschild radius. The Maxwell equations in this case are:
\numparts
\begin{eqnarray}
\frac{\rmi}{c}\partial_t{\mathcal F}^r=\partial_\theta\sin\theta{\left(1-\frac{r_s}{r}\right){\mathcal F}^\varphi}-\partial_\varphi\left(1-\frac{r_s}{r}\right)\frac{{\mathcal F}^\theta}{\sin\theta},\label{msa}\\
\frac{\rmi}{c}\partial_t{\mathcal F}^\theta=\partial_\varphi\frac{{{\mathcal F}^r}}{r^2\sin\theta}
-\partial_r\sin\theta\left(1-\frac{r_s}{r}\right){\mathcal F}^\varphi,\label{msb}\\
\frac{\rmi}{c}\partial_t{\mathcal F}^\varphi=\partial_r\left(1-\frac{r_s}{r}\right)\frac{{{\mathcal F}^\theta}}{\sin\theta}
-\partial_\theta\frac{{\mathcal F}^r}{r^2\sin\theta}.\label{msc}
\end{eqnarray}
\endnumparts
We shall seek the solutions of these equations in the spherical form (patterned after equations (\ref{mult1}) and (\ref{conn1})):
\begin{eqnarray}\label{mult2}
\fl\qquad\left[\begin{array}{c}
{\mathcal F}_r\\
{\mathcal F}_\theta\\
{\mathcal F}_\varphi
\end{array}\right]
=\left[\begin{array}{c}-l(l+1)r\sin\theta f_r(r)\\
-\lambda kr\,\partial_\varphi f_\varphi(r)
-\sin\theta\,\partial_\theta\partial_r r f_\theta(r)\\
\lambda kr\partial_\theta f_\varphi(r)
-1/\sin\theta\,\partial_\varphi\partial_r rf_\theta(r)
\end{array}\right]e^{-\rmi\lambda\omega t}Y_{lm}(\theta,\varphi).
\end{eqnarray}
After cumbersome calculations we obtain three equations for the radial functions $(f_r,f_\theta,f_\varphi)$,
\numparts
\begin{eqnarray}
\fl rf_r(r)-(r-r_s)f_\varphi(r)=0,\label{rada}\\
\fl \left[rf_\theta(r)\right]'
-\left[(r-r_s)f_\varphi(r)\right]'=0,\label{radb}\\
\fl r^2(r-r_s)f_\theta''(r)+(2r^2-rr_s)f_\theta'(r)+r_sf_\theta(r)
-l(l+1)rf_r(r)+k^2r^3f_\varphi(r)=0,\label{radc}
\end{eqnarray}
\endnumparts
where primes denote derivatives with respect to $r$. After the elimination of $f_\theta$ and $f_\varphi$, we obtain the following equation for the function $g=rf_r$:
\begin{eqnarray}
g''(r)+\frac{r_s}{r(r-r_s)}g'(r)+\left(\frac{k^2r^2}{(r-r_s)^2
}-\frac{l(l+1)}{r(r-r_s)}\right)=0.\label{rad}
\end{eqnarray}
This equation, known as the confluent Heun equation, does not have solutions in terms of known functions.

\begin{figure}
\centering
\includegraphics[width=9cm,height=5cm]{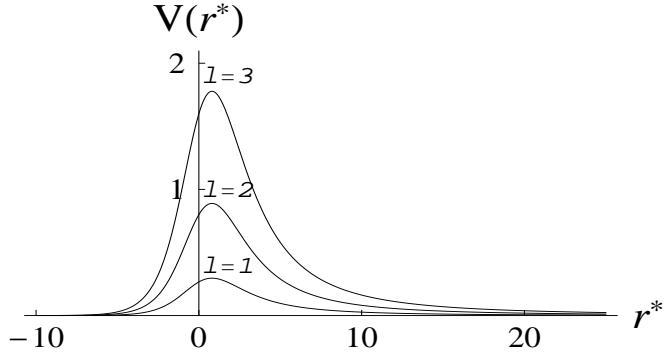}
\caption{Effective potential $V$ (measured in units of $r_s^{-2}$) as a\\function of the distance from the black hole (measured in units of $r_s$).\\The potential curves are drawn for three values of $l$.}\label{fig7}
\end{figure}

The physical meaning of the solutions may be best seen after the change of variables $r^*=r+r_s\ln(r/r_s-1)$. In the new radial coordinate $r^*$ the horizon, that lies at $r=r_s$ in the standard Schwarzschild coordinate, is pushed to $-\infty$. The function $r^*(r)$ cannot be explicitly inverted but $r(r^*)$ can be found numerically.

The equation in the variable $r^*$ has the form of the one-dimensional Schr\"odinger equation~\cite{mtw},
\begin{eqnarray}
-{\tilde g}''(r^*)+\frac{l(l+1)}{r^2}
\left(1-\frac{r_s}{r}\right){\tilde g}(r^*)=k^2{\tilde g}(r^*),\label{rad1}
\end{eqnarray}
where $r$ is to be treated as a function of $r^*$ and ${\tilde g}(r^*)=g(r(r^*))$. The second term on the left-hand side plays the role of the potential $V(r^*)$. The function $V(r^*)$ plotted in figure  \ref{fig7} has the form of a potential barrier. The maximum of the potential barrier has the value $4 l(l+1)/27$ and is located at $r^*=(3/2-\ln 2)r_s\approx 0.8 r_s$. Due to the repulsive character of the potential some part of the electromagnetic waves impinging on the black hole will bounce off; the rest will be sucked into the black hole. The reflection by the potential barrier is more pronounced for small values of $k$ and for large values of $l$. The insights drawn from the analogy with the Schr\"odinger equation are fully justified because the RS vector (represented here by ${\tilde g}(r^*)$) takes on complex values so that it can be treated as an analog of the Schr\"odinger wave function.

\section{The RS vector in a relativistic setting}\label{relco}

The three-dimensional RS vector is a part of the relativistic antisymmetric complex tensor $F^{\mu\nu}$,
\begin{eqnarray}
F^{\mu\nu}=\sqrt{-g}\left(g^{\mu\lambda}g^{\nu\rho}f_{\lambda\rho}
+\frac{\rmi}{2\sqrt{-g}}\epsilon^{\mu\nu\lambda\rho}
f_{\lambda\rho}\right),\label{rel1}
\end{eqnarray}
where $\epsilon^{\mu\nu\lambda\rho}$ is the Levi-Civita symbol (antisymmetric in all indices and $\epsilon^{0123}=1$). The Levi-Civita symbol by itself is not a tensor but a tensor density \cite{mtw}; it becomes a regular tensor after the division by $\sqrt{-g}$. Therefore, to be precise, $F^{\mu\nu}$ is the tensor density; it is a sum of two tensors multiplied by $\sqrt{-g}$. The definition of  $F^{\mu\nu}$ is valid in an arbitrary coordinate system, also in curved space.

Since the $f_{\lambda\rho}$ has 6 real components, only three components of the complex tensor $F^{\mu\nu}$ can be independent. Indeed, this tensor is self-dual; it transforms into itself under the following duality transformation:
\begin{eqnarray}
\frac{\rmi}{2\sqrt{-g}}\epsilon^{\mu\nu}_{\;\;\;\lambda\rho}
F^{\lambda\rho}=F^{\mu\nu}.\label{self}
\end{eqnarray}
The relativistic tensor $F^{\mu\nu}$ obeys Maxwell equations in the form:
\begin{eqnarray}
\partial_\mu F^{\mu\nu}=0.\label{maxcpl}
\end{eqnarray}
The real and the imaginary part of this complex equation reproduce the two pairs of Maxwell equations written in the relativistic notation,
\begin{eqnarray}
\partial_\mu\left(\sqrt{-g}f^{\mu\nu}\right)=0,\label{maxc1}\\
\partial_\mu f_{\nu\lambda}+\partial_\lambda f_{\mu\nu}+\partial_\nu f_{\lambda\mu}=0.\label{maxc2}
\end{eqnarray}

The three components of the RS vector can be identified with the space-time components of $F^{\mu\nu}$,
\begin{eqnarray}
{\bm F}=\left(F^{10},F^{20},F^{30}\right).\label{f2f}
\end{eqnarray}
However, due to the condition of self-duality (\ref{self}),
\begin{eqnarray}
\left(F^{23},F^{31},F^{12}\right)
=\frac{\rmi}{\sqrt{-g}}\left(F^{10},F^{20},F^{30}\right),\label{f2g}
\end{eqnarray}
the components of the RS vector can also be identified with the space-space components of $F^{\mu\nu}$. This explains why the components of the RS vector under Lorentz transformation transform among themselves (\ref{lor}), even though they are at the same time the components of a relativistic tensor $F^{\mu\nu}$ whose space-time components should mix with space-space components. In other words, the electric and the magnetic field vectors can be combined into two equivalent geometric objects: a complex vector ${\bm F}$ transforming under the complex orthogonal group $O(3,C)$ or a real second rank tensor $f_{\mu\nu}$ transforming under the real group $O(1,3)$.

In flat space and in Cartesian coordinates the Fourier representation (\ref{irep}) written in term of $F^{\mu\nu}$ has the following explicitly relativistic form \cite{qed}:
\begin{eqnarray}\label{irep2}
\fl\qquad F^{\mu\nu}(\bm r,t)=\frac{1}{(2\pi)^{3/2}}\int\!\frac{d^3k}{k}e^{\mu\nu}(\bm k)\left[f_+(\bm k)\rme^{\rmi\bm k\cdot\bm r-\rmi\omega t}+f_-^*(\bm k)\rme^{-\rmi\bm k\cdot\bm r+\rmi\omega t}\right].
\end{eqnarray}
The tensor $e^{\mu\nu}(\bm k)$ is self-dual; its components can be expressed in terms of the polarization vector ${\bm e}(\bm k)$.
\vspace{0.5cm}

\centerline{\Large{\bf Quantum theory}}
\addcontentsline{toc}{section}{\Large Quantum theory\hspace{20cm}}

\vspace{0.3cm}

\section{Canonical quantization of the electromagnetic field}

The standard procedure to formulate the quantum theory is the canonical quantization of the classical theory. This procedure was first applied to electromagnetism by Heisenberg and Pauli~\cite{hp}. Canonical quantization requires the identification of the canonical variables and this was properly done in the general case by Born and Infeld~\cite{bi}. The Maxwell equations clearly indicate the right identification: canonical variables are the field vectors $\bm D$ and  $\bm B$ whose time derivatives are determined by the evolution equations. It does not really matter whether one assigns the role of position to $\bm B$ and the role of momentum to $\bm D$ or vice versa. However, the first choice is more natural since the sources---the counterparts of the external mechanical forces---appear in the equation for $\bm D$. Either way, what is only needed are the Poisson brackets for canonical variables that will enable one to determine the commutators of the field operators. The basic requirement for the Poisson brackets is that they reproduce the equations of motion,
\begin{eqnarray}\label{eqm}
\partial_tQ(t)=\{Q(t),H\},
\end{eqnarray}
where $Q(t)$ is any variable and $H$ is the Hamiltonian of the system. The Poisson bracket for $\bm B$ and $\bm D$ in the form:
\begin{eqnarray}\label{pb}
\{B_i({\bm r},t),D_j({\bm r}',t)\}=\epsilon_{ijk}\partial_k\delta^{(3)}({\bm r}-{\bm r}')
\end{eqnarray}
not only reproduces the time dependence (\ref{eqm}), but it also assures the proper action of the remaining generators of the Poincar\'e group. Note that the Poisson brackets (\ref{pb}) are universal---they contain no material constants. This validates the choice of $\bm B$ and $\bm D$ (and not, for example, $\bm H$ and $\bm E$) as the canonical variables. According to the rules of canonical quantization, the Poisson brackets are replaced by the commutators of the corresponding operators (divided by $\rmi\hbar$). The commutation relations for ${\hat{\bm B}}$ and ${\hat{\bm D}}$ lead to the following commutators for the operators ${\hat{\bm F}}$ and ${\hat{\bm F}^\dagger}$ representing the RS vector and its Hermitian conjugate:
\begin{eqnarray}\label{qpb}
[{\hat F}_i({\bm r},t),{\hat F}^\dagger_j({\bm r}',t)]=\rmi\hbar c\epsilon_{ijk}\partial_k\delta({\bm r}-{\bm r}').
\end{eqnarray}
The photon picture emerges more directly when one uses the RS vector than in the standard formulation. After the Fourier transformation of the RS operators according to the formula (\ref{irep}) we arrive at the expression,
\begin{eqnarray}\label{irepq}
{\hat{\bm F}}(\bm r,t)=\int\!\frac{d^3k}{(2\pi)^{3/2}}{\bm e}(\bm k)\left[\hat{f}_+(\bm k)\rme^{\rmi\bm k\cdot\bm r-\rmi\omega t}+\hat{f}_-^\dagger(\bm k)\rme^{-\rmi\bm k\cdot\bm r+\rmi\omega t}\right],
\end{eqnarray}
where the classical amplitudes are replaced by the operators. In order to recover the commutation relations \eref{qpb} for the RS operator one must impose the following commutation relations on the operators $\hat{f}_\lambda$:
\begin{eqnarray}\label{crf}
[{\hat f}_\lambda(\bm k),{\hat f}_\lambda^\dagger(\bm k')]=\hbar \omega\,\delta_{\lambda\lambda'}\delta^{(3)}({\bm k}-{\bm k}').
\end{eqnarray}
Were it not for the factor $\hbar\omega$ these would be exactly the commonly used commutation relations for the photon annihilation and creation operators. We shall get rid of the factor $\hbar c$ by the following rescaling:
\begin{eqnarray}\label{cao}
{\hat f}_\lambda(\bm k)=\sqrt{\hbar c}\,{\hat a}_\lambda(\bm k),\;{\hat f}_\lambda^*(\bm k)=\sqrt{\hbar c}\,{\hat a}_\lambda^\dagger(\bm k).
\end{eqnarray}
The commutation relations for the annihilation and creation operators obtained after the rescaling are:
\begin{eqnarray}\label{cr}
[{\hat a}_\lambda(\bm k),{\hat a}_{\lambda'}^\dagger(\bm k')]=k\,\delta_{\lambda\lambda'}\delta^{(3)}({\bm k}-{\bm k}').
\end{eqnarray}
We retained the factor of $k$ in these relations because its presence is connected with the relativistic normalization of the photon wave functions, as will be shown in Sec.~\ref{mrep}. The formulas \eref{cao} justify our notation in \eref{dec} where we introduced complex conjugation in the second term of the Fourier transform. Finally, the Fourier representation of the RS field operator takes on the following form:
\begin{eqnarray}\label{irepq1}
\fl\qquad\quad{\hat{\bm F}}(\bm r,t)=\sqrt{\hbar c}\int\!\frac{d^3k}{(2\pi)^{3/2}}\,{\bm e}(\bm k)\left[a_+(\bm k)\rme^{\rmi\bm k\cdot\bm r-\rmi\omega t}+a_-^\dagger(\bm k)\rme^{-\rmi\bm k\cdot\bm r+\rmi\omega t}\right].
\end{eqnarray}
The time dependence of the annihilation and creation operators is given by $\exp(-\rmi\omega t)$ and $\exp(\rmi\omega t)$, as it should be. Quite often this time dependence is used to distinguish the annihilation from the creation operators. However, this method sometimes fails while the assignment based on the canonical commutation relations always works \cite{bbr}.

The generators of the Poincar\'e transformations in quantum electrodynamics are obtained by replacing the classical RS vectors in (\ref{gena}--d) by quantum field operators. Expressed in terms of the creation and annihilation operators, the generators have the form:
\numparts
\begin{eqnarray}
{\hat H}=\sum_\lambda\int\!\frac{d^3k}{k}\,a_\lambda^\dagger({\bm k})\,\hbar\omega\,a_\lambda({\bm k}),\label{genfa}\\
{\hat{\bm P}}=\sum_\lambda\int\!\frac{d^3k}{k}\,a_\lambda^\dagger({\bm k})\,\hbar{\bm k}\,a_\lambda({\bm k}),\\
\hat{{\bm M}}=\sum_\lambda\int\!\frac{d^3k}{k}\,a_\lambda^\dagger({\bm k})\left[\rmi\hbar{\bm D}_\lambda\times{\bm k}+\lambda\hbar{\bm n}\right]a_\lambda({\bm k}),\label{genfc}\\
\hat{{\bm N}}=\sum_\lambda\int\!\frac{d^3k}{k}\,a_\lambda^\dagger({\bm k})\,{\rmi}\hbar\omega{\bm D}_\lambda\,a_\lambda({\bm k}).\label{genfd}
\end{eqnarray}
\endnumparts
The normal ordering of the creation and annihilation operators removes the unphysical vacuum contributions.

\section{Quantum mechanics of photons}

Quantum theory of the electromagnetic field was obtained in the previous section directly from the classical Maxwell theory and not by the second quantization of quantum mechanics of photons. Usually, second quantization upgrades the wave function to the status of the field operator. However, the photon wave function is a controversial notion because it is impossible to endow it with {\em all} properties of wave functions for massive particles. Nevertheless it is possible to derive unambiguously quantum mechanics of photons by ``reverse engineering'' from quantum field theory. The RS vector is very useful in this reconstruction. To proceed, we consider the one-photon sector of Hilbert space and we will define the action of the important operators in this sector.

\subsection{Photon wave function in momentum representation} \label{mrep}

The most general one-photon state $|f\rangle$ is generated from the vacuum state $|0\rangle$ by the action of a superposition $a_f^\dagger$ of creation operators,
\begin{eqnarray}\label{1phot}
|f\rangle=a_f^\dagger\,|0\rangle=\sum_\lambda\int\!\frac{d^3k}{k}\,f_\lambda(\bm k)a_\lambda^\dagger(\bm k)|0\rangle.
\end{eqnarray}
The normalization of the state vector $|f\rangle$ translates into the following normalization condition for the functions $f_\lambda(\bm k)$:
\begin{eqnarray}\label{norm}
||f||^2=\sum_\lambda\int\!\frac{d^3k}{k}\,f_\lambda^*({\bm k})f_\lambda({\bm k})=1.
\end{eqnarray}
We used the same symbols $f_\lambda(\bm k)$ as in the Fourier expansion of the RS vector (\ref{irep}) because the two-component function $f_\lambda(\bm k)$ determines the state of the photon and fully deserves the name of the photon wave functions in momentum representation.

The existence of the photon wave function in {\em momentum representation} is a direct consequence of the theory of representations of the Poincar\'e group and its subgroup: the Lorentz group. According to this theory, fully developed by Wigner \cite{wig}, there are two one-dimensional representations $f_\pm(\bm k)$ corresponding to two circular photon polarizations. Under a Poincar\'e transformation $\Pi$ the transformed functions $'f_\pm$ are multiplied by ${\bm k}$-dependent phase factors,
\begin{eqnarray}\label{ptrans}
'f_\pm(\Pi\bm k)=\exp[\pm\rmi\phi(\Pi,\bm k)]f_\pm(\bm k).
\end{eqnarray}
These functions should be identified with the Fourier amplitudes of the RS vector appearing in \eref{irep} because they have identical transformation properties under the Poincar\'e group. However, we shall not prove it here since to prove that the transformation properties of the electromagnetic field under the Poincar\'e group induce appropriate phase transformations of the Fourier amplitudes is a rather cumbersome task. The transformation properties of the annihilation and creation operators must be the same as those of the corresponding functions $f$ since the RS operator must transform in the same way as its classical counterpart.

The expectation values ($\langle f|{\hat H}|f\rangle$, etc.) of the quantum generators (\ref{genfa}--d) in one-photon states coincide with the classical expressions (\ref{genca}--d). At this point we would like to comment on the significance of the factor $1/k$ in the volume element in \eref{1phot}. This factor produces an invariant volume element on the light cone and the whole formulation is relativistically covariant. The expression $f_\lambda^*({\bm k})f_\lambda({\bm k})\,d^3k/k$ is the invariant probability to find the photon with momentum $\hbar\bm k$ and polarization $\lambda$ in the volume $d^3k$ provided the function $f_{\lambda}(\bm k)$ is normalized. The normalization condition (\ref{norm}) is relativistically invariant, as it must be.

The action of the generators of the Poincar\'e transformations (\ref{genfa}--d) in quantum electrodynamics on the state vector $|f\rangle$ is equivalent to the following action on the functions $f_{\lambda}(\bm k)$:
\numparts
\begin{eqnarray}
{\hat H}f_{\lambda}(\bm k)&=\hbar\omega\,f_{\lambda}(\bm k),\label{gensa}\\
{\hat{\bm P}}f_{\lambda}(\bm k)&=\hbar{\bm k}\,f_{\lambda}(\bm k),\label{gensb}\\
{\hat{\bm M}}f_{\lambda}(\bm k)&=\hbar\left({\bm k}\times\frac{1}{i}\bm D_\lambda+\lambda{\bm n}\right)f_{\lambda}(\bm k),\label{gensc}\\
{\hat{\bm N}}f_{\lambda}(\bm k)&=\hbar\omega\,i\bm D_\lambda\,f_\lambda(\bm k).\label{gensd}
\end{eqnarray}
\endnumparts
We would have obtained the same results from the Wigner theory of representations. This makes the whole scheme fully consistent and one may identify the amplitudes of one-photon states (\ref{1phot}) with the photon wave functions in momentum space. Note that the quantum theory of the electromagnetic field could be, in principle, obtained by second quantization of the quantum mechanics of photons, i.e. by upgrading the photon wave functions in momentum space $f_\pm$ to the status of the annihilation operators.

\subsection{Photon wave function in coordinate representation} \label{crep}

The definition of the photon wave function in coordinate representation has well known difficulties. The straightforward Fourier transformation of the momentum wave function does not produce a field that transforms locally under the action of the Poincar\'e group. In order to identify the mathematical function that may serve as the photon wave function in the coordinate representation, one has to decide which properties one wants to keep and which properties one is willing to abandon. In our opinion a natural candidate is the RS vector. This decision is supported by the analogies between the description of the electromagnetic field in terms of the RS vector and quantum mechanics.

Of course, Silberstein writing his paper twenty years before the appearance of the wave function, could not have noticed the mathematical analogy with quantum mechanics, although his vector has many features of the quantum-mechanical wave function. First of all, the RS vector satisfies the evolution equation (\ref{maxa}) that has the form of the Schr\"odinger equation with the Hamiltonian ${\hat H}={\bm p}\!\cdot\!{\bm S}$,
\begin{eqnarray}\label{hsp}
\rmi\hbar\partial_t{\bm F}({\bm r},t)=\frac{\hbar}{\rmi}{\bm\nabla}\!\cdot\!{\bm S}\,{\bm F}({\bm r},t),
\end{eqnarray}
where ${\bm S}$ are the spin-one counterparts of the Pauli matrices,
\begin{eqnarray}\label{spin}
\fl\qquad S_x=\left[
\begin{array}{ccc}
0&0&0\\
0&0&-\rmi\\
0&\rmi&0
\end{array}
\right],\quad
S_y=\left[
\begin{array}{ccc}
0&0&\rmi\\
0&0&0\\
-\rmi&0&0
\end{array}
\right],\quad
S_z=\left[
\begin{array}{ccc}
0&-\rmi&0\\
\rmi&0&0\\
0&0&0
\end{array}
\right].
\end{eqnarray}

The analogy between the scalar product in quantum mechanics and the expression
\begin{eqnarray}\label{sp}
\langle{\bm F}_1|{\bm F}_2\rangle=\int\!d^3r\,{\bm F}^*_1(\bm r)\!\cdot\!{\bm F}_2(\bm r)
\end{eqnarray}
enables one to introduce such notions as the {\em complete set of solutions} and {\em change of basis}. The most important change of basis is the Fourier transformation. When such a transformation is applied separately to the real electric and magnetic field vectors, the Fourier components are not independent and one has to impose cumbersome reality conditions. This problem does not exist for the RS vector. The only constraint that has to be imposed is the transversality of the Fourier transform: ${\bm k}\!\cdot\!{\bm F}({\bm k})=0$.

The analogy between the probability density $\psi^*\psi$ and ${\bm F}^*\cdot{\bm F}$ can also be illustrated with the phenomena of wave packet spreading during the free time evolution. It follows from the standard Schr\"odinger equation that the dispersion of the probability distribution $\sigma_r^2=\langle({\bm r}-\langle{\bm r}\rangle)^2\rangle$ has the following time dependence \cite{bck}:
\begin{eqnarray}\label{tevol}
\sigma_r^2=\sigma_r^2|_{t=t_1}+(t-t_1)^2\sigma_p^2/m^2,
\end{eqnarray}
where $\sigma_p^2$ is the dispersion in momentum which for the free evolution is a constant of motion and $t_1$ is the time at which the wave packet attains its minimal size. An analogous formula holds in electromagnetism since the second derivative of the dispersion is proportional to the total energy. It follows from the Maxwell equation (\ref{maxa}) that
\begin{eqnarray}\label{tevol1}
\fl\qquad\frac{d^2}{dt^2}\int\!d^3r\,({\bm r}-\langle{\bm r}(t)\rangle)^2{\bm F}^*({\bm r},t)\cdot{\bm F}({\bm r},t)=c^2\int\!d^3r\,{\bm F}^*({\bm r},t)\cdot{\bm F}({\bm r},t).
\end{eqnarray}
This formula shows a quadratic time dependence because the total energy is a constant of motion. Thus, as in quantum mechanics, one obtains:
\begin{eqnarray}\label{tevol2}
&\int\!d^3r\,({\bm r}-\langle{\bm r}(t)\rangle)^2{\bm F}^*({\bm r},t)\cdot{\bm F}({\bm r},t)\nonumber\\
&\fl=\int\!d^3r\,({\bm r}-\langle{\bm r}(t_1)\rangle)^2{\bm F}^*({\bm r},t_1)\cdot{\bm F}({\bm r},t_1)+(t-t_1)^2c^2\int\!d^3r\,{\bm F}^*({\bm r},t)\cdot{\bm F}({\bm r},t),
\end{eqnarray}
where $t_1$ is the time at which the dispersion attains its lowest value.

In spite of these analogies there are important differences between the Schr\"odinger wave function $\psi$ and the RS vector. Even at first glance one can see that they differ in having different dimensions: ${\bm F}^*({\bm r},t)\cdot{\bm F}({\bm r},t)$ is the energy density while $\psi^*({\bm r},t)\psi({\bm r},t)$ is the probability density to find the particle at the position $\bm r$.

A more important difference stems from the nonexistence of a bona fide photon position operator. In particular, a natural choice---an operator $\hat{\bm r}$ whose action on ${\bm F}({\bm r},t)$ results in the multiplication by $\bm r$ as in quantum mechanics---is ruled out since the multiplication by $\bm r$ produces a function that is not divergenceless: ${\bm\nabla}\cdot x_i{\bm F}({\bm r},t)=F_i({\bm r},t)$. The nonexistence of the position operator precludes the existence of the photon probability density. Nevertheless, one may formulate the uncertainty relation for photons using the energy density.

\subsection{Uncertainty relation for photons}\label{ur}

In most general terms, the uncertainty relation for position and momentum imposes limitations on the spread of quantum systems both in the coordinate space and in the momentum space. In its standard form, this uncertainty relation involves the {\em dispersions} in position and momentum, but other measures of the spread (for example, entropic measures \cite{rbb}) have also been successfully used. We proposed in \cite{bb8} to base the uncertainty in position for photons on the energy density---the modulus squared of the RS vector. More precisely, we introduced the second moment of the energy distribution as a measure of the spread of the photon extension in coordinate space. The corresponding quantity is denoted by $\Delta r$ (and not $\Delta r^2$) because it has the dimension of length,
\begin{eqnarray}\label{delr}
\Delta r =\frac{1}{\hbar c||f||^2}\int\!d^3r\,{\bm r}^2{\bm F}^*(\bm r,t)\!\cdot\!{\bm F}(\bm r,t).
\end{eqnarray}
Division by the norm is necessary to make $\Delta r$ independent of the overall amplitude of the RS vector. In other words, $\Delta r$ is normalized per one photon. A natural partner of $\Delta r$ in the uncertainty relation is $\Delta p$,
\begin{eqnarray}\label{delp}
\Delta p =\frac{1}{||f||^2}\sum_{\lambda}\int\!\frac{d^3k}{k}\,\hbar \,|{\bm k}-\langle{\bm k}\rangle|\;|f_\lambda(\bm k)|^2,
\end{eqnarray}
that measures the spread of the photon wave function in momentum space. The lower bound for $\Delta r\Delta p$ was evaluated analytically in \cite{bb8} in the center of momentum frame ($\langle{\bm k}\rangle=0$) and it gives the uncertainty relation in the form:
\begin{eqnarray}\label{ur1}
\Delta r\Delta p\ge 4\hbar.
\end{eqnarray}
The lower bound is much larger then $3/2\hbar$ appearing in the three-dimensional Heisenberg uncertainty relation in nonrelativistic quantum mechanics. This is due to the fact that photon owing to its relativistic nature hinders localization. The photon state that saturates this inequality is the Synge solution of Maxwell equations \cite{bb8} described in Sec.~\ref{assort}.

\section{The RS vector and the spinor calculus}

In order to underscore the interpretation of the RS vector as the photon wave function we shall now prove that it fits very well into the general scheme of wave functions for massless relativistic particles. Relativistic properties of fields and particles find their apt description in terms of spinors and the RS vector does not depart from this general rule. The RS vector is directly related to the second rank relativistic symmetric spinor field $\phi_{AB}(\bm r,t)$. Such a spinor field has only three independent components $(\phi_{00},\phi_{11},\phi_{01})$. Assuming the relations between $\phi_{AB}$ and the RS vector in the following form:
\begin{eqnarray}\label{f2p}
F_x=\phi_{11}-\phi_{00},\quad F_y=-\rmi(\phi_{11}+\phi_{00}),\quad
F_z=2\phi_{01}=2\phi_{10},
\end{eqnarray}
we can prove the equivalence of the spinorial equations:
\begin{eqnarray}\label{max2spn}
\partial_t\phi_{AB}
=-c\left({\bm\sigma}\cdot{\bm\nabla}\right)_A^{\;\;C}\phi_{CB},
\end{eqnarray}
with the Maxwell equations. To prove this equivalence we rewrite (\ref{max2spn}) in full matrix form:
\begin{eqnarray}\label{max2spn1}
\fl\qquad\partial_t\left[\begin{array}{cc}
\phi_{00}&\phi_{01}\\
\phi_{01}&\phi_{11}
\end{array}\right]
=-c\left[\begin{array}{cc}
\partial_z\phi_{00}+(\partial_x-\rmi\partial_y)\phi_{01}&
\partial_z\phi_{01}+(\partial_x-\rmi\partial_y)\phi_{11}\\
(\partial_x+\rmi\partial_y)\phi_{00}-\partial_z\phi_{01}&
(\partial_x+\rmi\partial_y)\phi_{01}-\partial_z\phi_{11}
\end{array}\right],
\end{eqnarray}
where we employed the standard representation of Pauli matrices. Next, we use the relations (\ref{f2p}) to reproduce the Maxwell equations (\ref{maxa}). The symmetry of the matrix on the left-hand side of (\ref{max2spn1}) imposes the condition that the right-hand side is also symmetric, i.e,
\begin{eqnarray}\label{sym}
\partial_z\phi_{01}+(\partial_x-\rmi\partial_y)\phi_{11}=
(\partial_x+\rmi\partial_y)\phi_{00}-\partial_z\phi_{01}.
\end{eqnarray}
This equality implies that the divergence of the RS vector vanishes. Thus, there is a full equivalence of the description of the electromagnetic field in terms of the RS vector and in terms of the symmetric second rank spinor. However, when it comes to concrete applications the use of the RS vector is much more convenient due to its direct connection with the electric and magnetic field vectors.

There is a simple connection between the spinor $\phi_{AB}$ and the complex tensor $F^{\mu\nu}$ defined in (\ref{rel1}) that can be expressed in terms of Pauli matrices combined into the spin-tensor $S^{\mu\nu AB}$,
\begin{eqnarray}\label{st}
S^{0k}=-S^{k0}=-\rmi\sigma_2\sigma_k,\quad
S^{ij}=-S^{ji}=\epsilon^{ijk}\sigma_2\sigma_k,
\end{eqnarray}
where we suppressed the $AB$ indices. With the use of (\ref{f2p}) we obtain:
\begin{eqnarray}\label{st1}
F^{\mu\nu}=S^{\mu\nu AB}\phi_{AB}.
\end{eqnarray}
Null electromagnetic waves have a very simple form in terms of spinors. Namely, they are described by just one first rank spinor field $\phi_A$ ($\phi_{AB}=\phi_A\phi_B$). Of course, not every spinor field $\phi_A$ gives rise to a null solution of Maxwell equations. The second rank spinor $\phi_{AB}$ must still satisfy (\ref{max2spn}).

The equation (\ref{max2spn}) is a special case of the general wave equation for a massless particle. Regardless of the spin value $s$, these equations have always the same general form.
\begin{eqnarray}\label{genspn}
\partial_t\phi_{AB\dots W}
=-c\left({\bm\sigma}\cdot{\bm\nabla}\right)_A^{\;\;Z}\phi_{ZB\dots W}.
\end{eqnarray}
They differ only in the rank of the spinor that is equal to $2s$. The first-rank spinor describing a spin 1/2 massless neutrino satisfies the Weyl equation
\begin{eqnarray}\label{weyl}
\partial_t\phi_{A}
=-c\left({\bm\sigma}\cdot{\bm\nabla}\right)_A^{\;\;Z}\phi_{Z}.
\end{eqnarray}
The fourth-rank spinor satisfies the equation:
\begin{eqnarray}\label{grav}
\partial_t\phi_{ABCD}
=-c\left({\bm\sigma}\cdot{\bm\nabla}\right)_A^{\;\;Z}\phi_{ZBCD}.
\end{eqnarray}
It describes gravitons---the quanta of the gravitational field.

\section{Summary}

In this review we have shown that the RS vector is an ideal tool to unify the classical Maxwell theory and the quantum mechanics of photons. On the one hand the RS vector gives a concise description of the classical electromagnetic field. On the other hand it is the best possible photon wave function \cite{pwf}. In the study of solutions of the Maxwell equations complex functions to describe real fields are widely used since ``Exponentials are much easier to manipulate than sines and cosines'' \cite{griff}. However, in the standard approach one only keeps the real part as physically significant. The RS vector offers all advantages of working with complex functions but there is no redundance in this description; the real part and the imaginary part are both physical. Nevertheless, if the role of the RS vector were restricted only to classical theory, we would have to agree with Stratton \cite{stratton} who in his monograph wrote: ``It has been shown by Silberstein, Bateman, and others that the equations satisfied by the fields and potentials may be reduced to a particularly compact form by the construction of a complex vector whose real and imaginary parts are formed from the vectors defining the magnetic and electric fields. The procedure has no apparent physical significance but frequently facilitates analysis''. In our review we have argued that it is the role of the RS vector as the photon wave function that brings forth physical significance denied by Stratton who was not aware of a dual role of the RS vector. The use of the RS vector as a photon wave function enables one to incorporate the quantum mechanics of photons into the general scheme valid for all relativistic particles.

\ack

We would like to thank {\L}ukasz Rudnicki for helpful comments. This research was partly supported by the grant from the Polish Ministry of Science and Higher Education for the years 2010--2012.

\end{document}